# High Open Circuit Voltages in *pin*-Type Perovskite Solar Cells through Strontium Addition

Pietro Caprioglio □†, Fengshuo Zu ǂ‡, Christian M. Wolff □, José A. Márquez Prietoǂ, Martin Stolterfoht □, Norbert Koch ǂ‡, Thomas Unoldǂ, Bernd Rech §, Steve Albrecht †§1 and Dieter Neher □1

□ University of Potsdam, Institut für Physik und Astronomie, Potsdam, Germany

‡ Humboldt-Universität, Institut für Physik, Berlin, Germany

§ Helmholtz-Zentrum Berlin, Institute for Silicon Photovoltaics, Berlin, Germany

†Helmholtz-Zentrum Berlin, Young Investigator Group Perovskite Tandem Solar Cells, Berlin, Germany

ǂHelmholtz-Zentrum Berlin für Materialien und Energie GmbH, Berlin, Germany

## Abstract

The incorporation of even small amounts of strontium (Sr) into lead-based quadruple cation hybrid perovskite solar cells results in a systematic increase of the open circuit voltage ($V_{oc}$) in *pin*-type perovskite solar cells. We demonstrate via transient and absolute photoluminescence (PL) experiments how the incorporation of Sr significantly reduces the non-radiative recombination losses in the neat perovskite layer and specifically at the perovskite/$C_{60}$ interface. We show that Sr segregates at the perovskite surface, where it induces important changes of morphology and energetics. Notably, the Sr-enriched surface exhibits a wider band gap and a more *n*-type character, accompanied with significantly stronger surface band bending. As a result, we observe a significant increase of the quasi-Fermi level splitting in the neat perovskite by reduced surface recombination and more importantly, a strong reduction of losses attributed to non-radiative recombination at the interface to the $C_{60}$ electron-transporting layer. The resulting solar cells exhibited a $V_{oc}$ of 1.18 V, which could be further improved to nearly 1.23 V through addition of a thin polymer interlayer, bringing the non-radiative voltage loss to only 110 meV. Our work shows that simply adding a small amount of Sr to the precursor solutions induces a beneficial surface modification in the perovskite, without requiring any post treatment, resulting in high efficiency solar cells with power conversion efficiency (PCE) up to 20.3%. Our results demonstrate very high $V_{oc}$ values and efficiencies in Sr-containing quadruple cation perovskite *pin* solar cells and highlight the imperative importance of addressing and minimizing the recombination losses at the interface between perovskite and charge transporting layer.

---

[1] corresponding author: neher@uni-potsdam.de, steve.albrecht@helmholtz-berlin.de

**Introduction**

Organic-inorganic halide perovskites are considered one of the most promising materials for photovoltaic applications due to their rather easy and low-cost fabrication, as well as outstanding optoelectronic properties. Notably, these semiconductors combine a high absorption coefficient with panchromatic absorption of light[1] with a long carrier diffusion length[2,3], allowing efficient photon absorption and charge extraction for a typical active layer thickness of only 500 nm. Another peculiarity of hybrid perovskites is that defects create mostly shallow energy levels, allowing high open circuit voltage ($V_{oc}$) and long carrier lifetime[4]. Moreover, perovskite materials can be obtained from in-nature abundant precursors, which potentially reduce further the costs of future large scale production. Despite the fact that the first full solid state perovskite solar cell was reported only in 2012, with a power conversion efficiency (PCE) of 9.7%[5], this technology has experienced a tremendous improvement[6–8], currently reaching a record PCE of 22.7% [9]. Regardless of the state of the art of perovskite solar cells and their astonishing performances, the metrics fill factor (*FF*) and $V_{oc}$ are currently still limiting their PCE. Thus, in order to achieve higher efficiencies the limiting agents of these parameters need to be understood and minimized[10,11]. It has been realized that non-radiative recombination at or across the interface between the perovskite absorber and the adjacent charge-transporting layer (CTL) constitutes a major recombination loss[12,13]. Motivated by these findings, several approaches such as surface treatment or passivation, interlayers and compositional engineering have been applied to boost the $V_{oc}$ of these devices[8,14–16]. However, most of these approaches require a multistep deposition process, where either the existing perovskite layer is altered through a post-deposition treatment or an additional thin layer is deposited on top. Both approaches are not suitable when aiming at fast production schemes. Alternatively, attempts have been made to increase the performance of perovskite devices by adding suitable components to the perovskite precursor solution[17–19]. Such additives were shown to enrich at grain-boundaries or at the perovskite surface, paving the way for suppressing unwanted non-radiative recombination while avoiding a multistep preparation scheme [e.g. TOPO[16], and some other small and large molecules]. One recent example of this strategy is the addition of $SrI_2$ to the precursor solution of hybrid and inorganic perovskites. It has been shown that $Sr^{+2}$ partially substitute $Pb^{2+}$ in the perovskite lattice [20,21], owing to the nearly almost identical ionic radii of both ions ($Sr^{2+}$ = 132 pm, $Pb^{2+}$ = 133 pm)[21]. Recent results suggested, however, that Sr segregates preferentially at the surface of solution-processed perovskites films, going along with specific changes of the photovoltaic parameters.

For example the addition of 1-2 % $SrI_2$ to the $MAPbI_3$ precursor solution increased the PCE, of an *pin*-type device from 12 % to nearly 15 %, mainly through an increase in $J_{sc}$ and the FF, while the $V_{oc}$ was actually reduced [22]. The overall improvement in device efficiency, was attributed to an increased carrier lifetime in combination with surface passivation, resulting in an improved charge extraction. More recently Lau *at al.* reported $Sr^{2+}$ insertion into a $CsPbI_3$ perovskite[23]. Here, addition of Sr at a concentration of 2 % improved all photovoltaic parameters of a *nip*-type solar cell, resulting in a stabilized PCE of nearly 11 %. Based on an improved PL lifetime and predominant Sr surface aggregation, the authors concluded that Sr mainly acts as a surface passivating agent. Furthermore, also in Cl-containing perovskite the presence of Sr has shown positive effects on the performance of actual devices[24,25]. On the other hand, a more recent paper demonstrated a significant reduction of all photovoltaic parameters of *nip*-type devices when adding Sr to a MAPI precursor solution[26].

Here, we apply this approach to efficient *pin* type devices comprising a solution processed quadruple cation perovskite[27] sandwiched between the hole-transporting polymer PTAA and a $C_{60}$ electron-transporting layer. These *pin* type cells attracted considerable attention as they require only ultrathin undoped charge transport layers (e.g. 8 nm PTAA, 30 nm $C_{60}$)[11] and moderate annealing temperatures. However, the efficiency of such devices was shown to be largely limited by the perovskite/$C_{60}$ interface, limiting the $V_{oc}$ to values of about 1.11 V for a perovskite with a band gap of around 1.62 eV bandgap, if no further treatments are applied[28]. We find that addition of Sr leads to a large reduction of non-radiative recombination loss in the device, with an increase of the $V_{oc}$ by 70 mV, up to a remarkable value of 1.18 V. A combination of methods, including transient and absolute photoluminescence (PL), second ion mass spectroscopy (SIMS), scanning electron microscopy (SEM), and photoelectron spectroscopy (PES) is applied to arrive at a comprehensive picture of the morphology and surface energetics of the neat perovskite layers, with and without Sr. Our study confirms that Sr segregates mostly at the perovskite surface, where it widens the band gap and induces a stronger *n*-type surface band bending. We demonstrate that these modifications limit the accessibility of the surface to photo-generated holes, thereby enhancing the quasi-Fermi level splitting and reducing interface-mediate non-radiative recombination with $C_{60}$ in the Sr-containing perovskite samples.

**Device Structure and Materials**

We employed solar cells with a simple "inverted" configuration where light enters through the hole-extracting contact (denoted here as *pin* geometry) [29], as shown in Fig 1 a). Acknowledging that, currently, best performing perovskite solar cells utilize a combination of inorganic and organic transport layers in a regular structure (*nip*), in this work we focus exclusively on undoped, fully organic transport layers. Inverted solar cells are prone to reduced hysteresis, low processing temperatures and recently gained strong interest in perovskite based multijunction devices[30]. Here, a very thin layer (<10 nm) of the transparent and semiconducting polymer poly[bis(4-phenyl)(2,4,6-trimethylphenyl)amine] (PTAA) serves as the hole-transporting material. On top of the hole contact, around 400 nm thickness of active material is deposited by spin-coating utilizing a solution of different quadruple cation perovskite compositions as shown in Figure 1b) [31,32]. To analyse the effect of Sr addition, the Pb is partially replaced by Sr varying from 0% to 2%, see below. As an electron transporting layer, 30 nm of $C_{60}$ is deposited on top of the perovskite through thermal evaporation, followed by the top contact constituted by 8 nm of bathocuproine (BCP) and 100 nm of copper (Cu). It has been shown before that $C_{60}$ does not form an ideal electron selective contact in combination with $MAPbI_3$, and that it mediates significant non-radiative recombination[10].

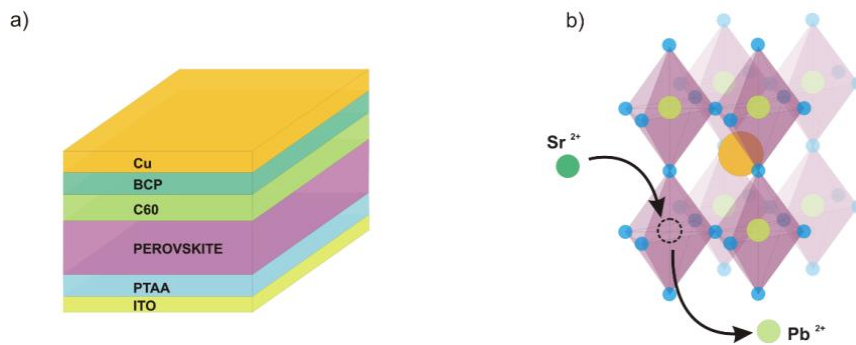

*Figure 1: a) Solar cell device structure representing the layer stack utilized here, b) $ABX_3$ perovskite cubic structure with the different species implemented in this study. The figure indicates how the $Sr^{2+}$ can take the place of $Pb^{2+}$ into the perovskite lattice.*

Our group has recently shown that this device design can yield high efficiencies even with undoped charge transporting layers[11]. For this work, the starting perovskite composition is a quadruple cation perovskite based on the mixture of lead iodide ($PbI_2$) and lead bromide ($PbBr_2$) with four cation salts, namely methylammonium bromide (MABr), formamidinium iodide (FAI), cesium iodide (CsI) and rubidium iodide (RbI), as previously reported by Saliba et al.[27] Here, the abovementioned compounds in the composition $(MA_{0.17}FA_{0.83})Pb(I_{0.83}Br_{0.17})_3$, were dissolved in dimethylformamide (DMF) and dimethyl sulfoxide (DMSO) in a ratio of 4:1 to which a solution of CsI and RbI, again in a DMF:DMSO (4:1) solvent mixture, was added in order to obtain a final molar composition of $(Rb_{0.05}(Cs_{0.05}(MA_{0.17}FA_{0.83})_{0.95})_{0.95})Pb(I_{0.83}Br_{0.17})_3$. Finally, a small amount of $SrI_2$ dissolved in DMF:DMSO (4:1) was added in different quantities to yield a Sr/Pb ratio of 0%, 0.05%, 0.1%, 0.3%, 0.5%, 1%, 2%, with the goal to study the effect of the Sr insertion from fairly small to considerably high quantities.

## Results and Discussions

### Optoelectronic Properties

Already a small quantity of Sr (0.05%) affects the photovoltaic performance, especially the $V_{oc}$, which increases from 1.10 V to 1.12 V. The rather low $V_{oc}$ of the Sr-free device is most likely caused by the poor selectivity of $C_{60}$ as an electron-extracting contact, as mentioned above. Further increasing the Sr/Pb ratio to 2% in various steps, a consistent increase of $V_{oc}$ from 1.12 V to 1.18 V is found, see Figure 2b), highlighting an exceptional $V_{oc}$ increase of 70 mV compared to the reference cell. A $V_{oc}$ of 1.18 V, for a perovskite absorber with a band gap of 1.6 eV in a *pin* solar cell architecture is only enabled by a strong minimization of the energy losses. The resulting cells show a power conversion efficiency of 20.3 % for the record device, well exceeding the value of any other Sr containing perovskite reported so far [22–25]. Higher Sr concentrations (Fig. S1B and Fig. S1D, SI) did not improve the $V_{oc}$ further, while causing a significant drop of the *FF* and $J_{sc}$. For this reason, we refer to a 2% Sr addition as the optimized amount and use this concentration for further comparisons. Adding Sr until 2% does not compromise the $J_{sc}$, but a systematic mild decrease of the average fill factor is observed (Table 1). Noteworthy, regardless the presence of Sr the devices here do not show any evidence of hysteresis effect during *J-V* scans (Fig. S1A, SI). Notably, an opposite effect has been reported when Sr was added to simple MAPbI$_3$ in a solution processed *pin* device, as described above[22]. Adding Sr to MAPbI$_3$ led to a marked increase in $J_{sc}$ and *FF*, assigned to improved charge extraction while the $V_{oc}$ decreased by 50 meV[22]. We cannot resolve this discrepancy but we note that our devices differ from those employed in Ref.32 substantially (the choice of the perovskite and all charge-transporting layers as well as a different way of perovskite layer preparation). Following our previous studies[10], we tried to further enhance the $V_{oc}$ by adding a thin (less then 5nm) insulating polystyrene (PS) layer at the interface between perovskite and C$_{60}$. The resulting solar cell shows an extraordinarily high $V_{oc}$ of nearly 1.23 V due to the significantly reduced interface recombination. This result highlights the $V_{oc}$ potential of well optimized electron selective contact. However, as reported in Fig. S1C, SI, the inclusion of the ultra-thin insulating PS interlayer lead to a significant reduction the *FF* by probably limiting the extraction of electrons from the perovskite layer. For a better comparison, a box charts representing all parameters for the most important type of devices analysed in this work is presented in Fig. S1D, SI. As our work is mostly concerned with the suppression of interfacial recombination though mixed solution processing, we will not consider this approach further in the work.

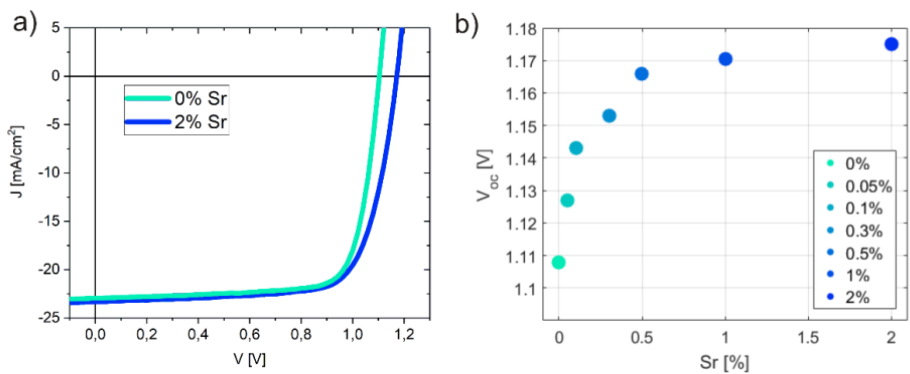

*Figure 2: a) J-V characteristics in reverse scan (0.1 V/s with voltage step of 0.02V) under simulated AM 1.5G illumination calibrated to 100 mW/cm² of the best devices for the two cases 0% Sr and 2% Sr. b) Averaged V$_{oc}$ values at 100 mW/cm² illumination as a function of Sr content based on 10 cells for each Sr concentration.*

*Table 1: Averaged J-V parameters for devices with different Sr concentrations including the standard errors based on a statistics of 10 cells for each Sr concentration. Values are taken from J-V scans with scan rate of 0.1 V/s and voltage step of 0.02V. As shown in Fig. S1 SI forward and reverse scan gives identical values showing total absence of hysteresis effect. The record parameters are reported in brackets.*

| Sr [%] | $V_{oc}$ [V] | $J_{sc}$ [mA/cm$^2$] | FF [%] | PCE [%] |
|---|---|---|---|---|
| 0 | 1.108 ±0.007 (1.121) | 22.0 ±0.6 (22.9) | 76.0 ±0.8 (77.6) | 19.2 ±0.3 (19.4) |
| 0.05 | 1.127 ±0.005 (1.133) | 21.4 ±0.5 (22.2) | 73.0 ±2.0 (75.8) | 18.4 ±0.6 (18.9) |
| 0.1 | 1.145 ±0.002 (1.150) | 22.0 ±0.2 (22.2) | 73.2 ±0.5 (74.0) | 18.8 ±0.1 (19.1) |
| 0.3 | 1.150 ±0.003 (1.153) | 21.4 ±0.3 (21.7) | 71.0 ±1.0 (72.6) | 18.2 ±0.4 (18.7) |
| 0.5 | 1.166 ±0.004 (1.169) | 21.9 ±0.1 (22.0) | 70.9 ±0.2 (71.1) | 18.3 ±0.6 (18.7) |
| 1 | 1.170 ±0.003 (1.174) | 21.4 ±0.4 (22.9) | 71.0 ±1.0 (72.5) | 18.8 ±0.4 (19.2) |
| 2 | 1.175 ±0.004 (1.180) | 22.6 ±0.4 (23.2) | 70.0 ±2.0 (74.0) | 19.7 ±0.7 (20.3) |
| 5 | 1.168 ±0.006 (1.178) | 15.0±0.5 (16.3) | 64.8±2.0 (66.8) | 12.3±0.6 (13.1) |

We note here that the addition of Sr causes (on average) in a slight increase of $J_{sc}$. We don't have a conclusive interpretation for this phenomenon currently at hand, but it might be related to the larger surface roughness of the Sr-containing perovskite. The higher $J_{sc}$ current is also reflected in the slightly larger external quantum efficiency (EQE$_{PV}$) (Fig. S2A SI). The $J_{sc}$ values calculated from the integrated EQE$_{PV}$ spectrum (Fig. S2A) match within a 5% deviation with the $J_{sc}$ measured in *J-V* scans. Importantly, the position and shape of the onset of the EQE$_{PV}$ (Fig. S2C, SI) remain almost unaltered within sample reproducibility on Sr addition in the range up to 2%. A widening of the band gap, which might be assumed as a cause of the $V_{oc}$ increase after composition modification[27,33,34], is not found here. This is also supported by absorption measurements (Fig. S2B, SI) which show no shift in the absorption onset when changing the composition from 0% Sr to 2% Sr. Overall, the data show that Sr addition has no effect on the bulk energy gap, consistent with what was reported previously[18,22], meaning that the observed increase in $V_{oc}$ must be related to reduced recombination.

## Time-Resolved and Steady-State Photoluminescence

Time resolved photoluminescence lifetime (TRPL) measurements were applied to investigate the effect of Sr addition on the charge carrier recombination in the bare perovskite material The TRPL traces for neat perovskite films on glass substrates containing different amounts of Sr, shown in Figure 3 a), are in accordance with a double exponential decay model [35]:

$$I_{PL}(t) = A \cdot e^{\left(-\frac{t}{\tau_{fast}}\right)} + B \cdot e^{\left(-\frac{t}{\tau_{slow}}\right)}$$

(1)

In equation 1, the parameters A and B are the relative amplitudes for the fast and slow lifetimes, $\tau_{fast}$ and $\tau_{slow}$, respectively.

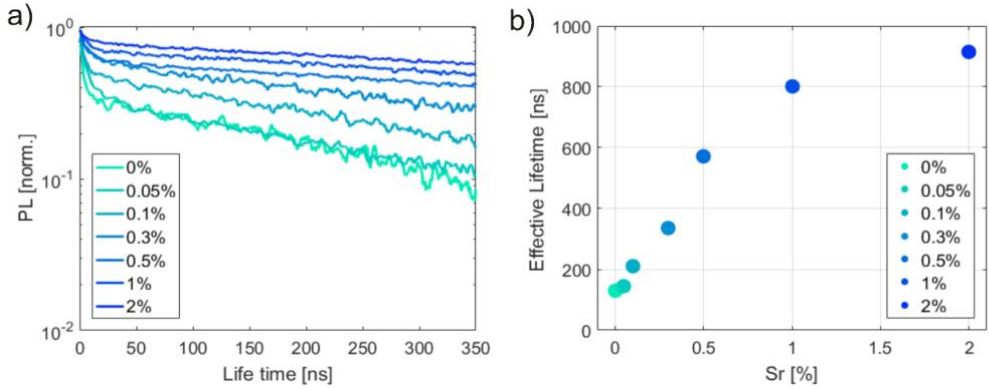

*Figure 3: a) Time resolved photoluminescence (TRPL) traces of neat perovskite layers for the different Sr concentrations measured in inert atmosphere, normalized to the initial transient peak. b) Calculated effective life times from fitting of the TRPL traces according to equation 1 and 2.*

Commonly, the initial fast decay (below 20 ns) is associated to the capture of charges by trap states and recombination at grain boundaries or at the surface[36]. An increase of the fast photoluminescence lifetime, indicated by the first initial decay in Fig. 3 a), point out that the addition of Sr either reduces the concentration of such traps acting as a passivating agent, or limits the accessibility of these traps to photogenerated free carriers. At the same time, TRPL traces also show an improvement of the slow (longer than 50 ns) photoluminescence lifetime, suggesting that non-radiative recombination, commonly associated with the slow decay and most likely due to processes happening at the surface[37], is considerably attenuated. An effective (amplitude averaged) photoluminescence lifetime $\tau_{eff}$ was then calculated according to the following equation[38]:

$$\tau_{eff} = \frac{A \cdot \tau_{fast} + B \cdot \tau_{slow}}{A + B}$$

(2)

The effective lifetimes are plotted in Fig. 3b) for the different compositions. The addition of Sr has a significant effect on the PL lifetime. With the addition of 2% Sr an extraordinarily long PL lifetime of almost 1 µs is measured. The positive effect of Sr addition on the PL lifetime found here, is qualitatively in agreement with the enhancement of carrier lifetime after insertion of Sr in MAPbI3 as deduced from transient microwave conductivity experiments[22] or with the increase of the PL lifetime in perovskites of different compositions[23,24]. Given the fact that the addition of Sr increases both PL lifetime and $V_{oc}$, but it has basically no effect on the shape and amplitude of the EQE or absorption spectrum, we propose that Sr mainly reduces the strength of non-radiative recombination. In order to quantify this reduction of energy losses in the bare absorber, we measured the quasi-Fermi level splitting (QFLS) by means of the absolute PL intensity measurements. Here, we make use of Würfel's generalized Plank's law[39] describing the non-thermal emission of a semiconductor:

$$I_{PL}(E) = \frac{2\pi E^2 a(E)}{h^3 c^2} \frac{1}{\exp\left(\frac{E - \Delta E_F}{kT}\right) - 1}$$

(3)

Here, $h$ is Planck's constant, $c$ is the speed of light, $E$ is the photon energy, $\Delta E_F$ the QFLS, $a(E)$ is the spectral absorptivity, and $k$ is the Boltzmann constant. Approximating the Bose-Einstein distribution with a Boltzmann distribution ($E - \Delta E_F \gg kT$) and assuming the absorptivity $a(E)=1$ for energies above the band gap, the PL intensity $I_{PL}$ can be expressed as function of $E$ where the QFLS, $\Delta E_F$, can be extrapolated from a fit of $\ln(I_{PL}/E^2)$ above the band gap [38]:

$$\ln\left(\frac{h^3 c^2 I_{PL}(E)}{2\pi E^2}\right) = \frac{\Delta E_F}{kT} - \frac{E}{kT}$$

(4)

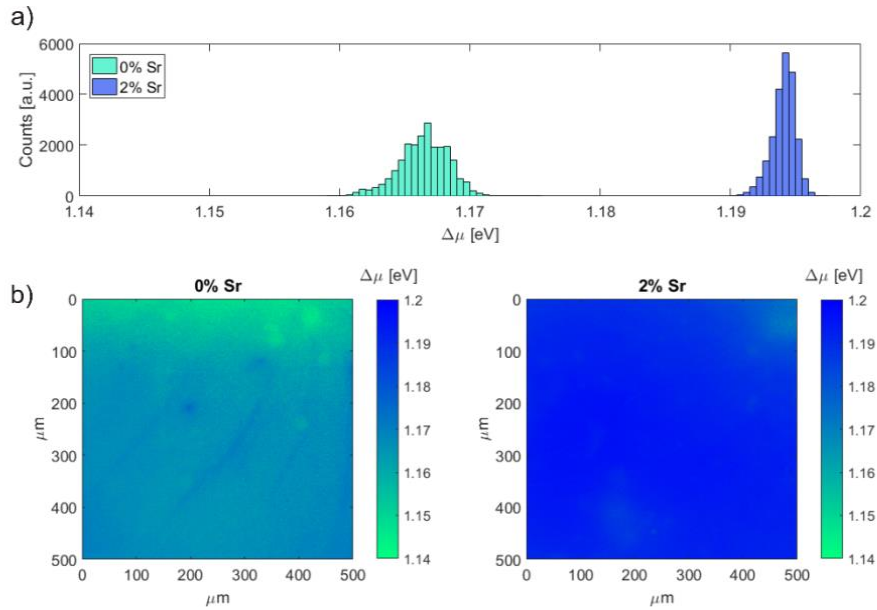

*Figure 4: a) Histogram of the quasi-Fermi level splitting (QFLS) extrapolated from the PL maps (0.5 cm$^2$) of a neat perovskite layer with 0% and 2% Sr represented in b). The maps represent the QFLS distribution on the surface of the layer in a range from 1.14 eV to 1.2 eV. Errors here are considered as the standard deviation of the respective histograms.*

Fig. 4 b) shows the QFLS maps calculated from the PL intensity mapping using Eq. 4 for two samples of neat perovskite with 0% and 2% Sr. In Fig. 4 a) the histograms for the corresponding QFLS distributions are reported. In the case of 2% Sr a high QFLS of $\Delta E_F = 1.194 \pm 0.001$ eV (FWHM =

2.3meV) was extrapolated from absolute PL mapping, whereas in the case of 0% Sr the quasi Fermi levels splitting was only $\Delta E_F = 1.167 \pm 0.002$ eV (FWHM = 5meV), indicating that the losses correlated to non-radiative recombination are considerably lower and the maximum achievable $V_{oc}$ higher when Sr is incorporated. This finding is consistent with the observed increase of the PL lifetimes. Also, the QFLS distribution of the Sr-containing device is significantly narrower, indicating that the Sr-addition also reduces the spatial inhomogeneity of recombination pathways across the perovskite film.

**Electroluminescence Efficiency**

A simple method to investigate radiative and non-radiative recombination in a full device is to measure the external electroluminescence efficiency ($EQE_{EL}$) by applying a forward bias to the solar cell in the dark operating it as a light-emitting diode (LED). Fig. 5 displays $EQE_{EL}$ for the 0% and the 2% samples for a range of applied voltages around the $V_{oc}$ and the corresponding injected dark currents.

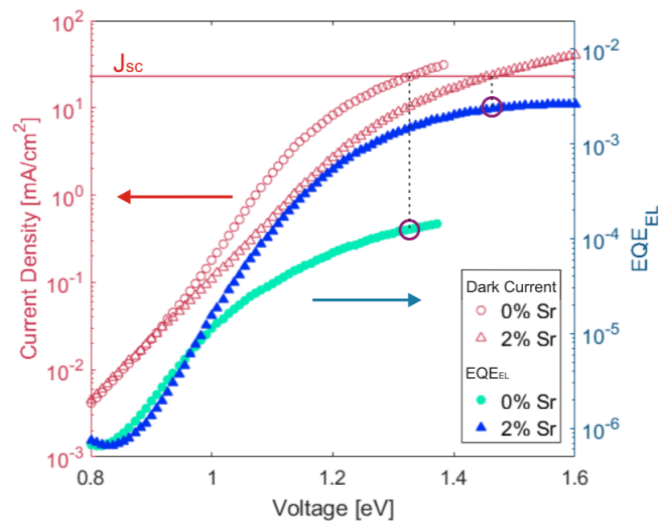

*Figure 5: The external electroluminescence efficiency $EQE_{EL}$ (right y-axis) and the injected dark current (left y-axis) for complete solar cells without (0% Sr) and with 2% of Sr as a function of the applied voltages. Dashed*

*lines and circles indicate the applied voltage, where the dark injected current is equal to the $J_{sc}$ under simulated AM1.5G illumination.*

When Sr is added, the EL efficiency is increased by more than one order of magnitude, reaching a remarkably high $EQE_{EL}$ value of $2.5 \cdot 10^{-3}$ for injection currents approaching $J_{sc}$ under illumination, denoting stronger emissive behaviour and reduction of non radiative recombination also in the complete device structure. A luminescence efficiency of 0.25% is among the best values reported in literature so far[10,40]. $EQE_{EL}$ values are taken for an injected current density equal to $J_{sc}$ (being a good approximation of the generation current $J_G$), meaning that the so-determined $EQE_{EL}$ is a good measure for the radiative recombination efficiency at $V_{oc}$ under 1 sun illumination and, therefore, relevant for the calculation of the radiative and non-radiative losses presented here. Following the approach from Rau[41] we calculated exceptionally low non-radiative losses of $\Delta V_{oc,non-rad} = \frac{k_B T}{e} ln\left(\frac{1}{EQE_{EL}}\right) = 0.161\ eV$ at $T = 300\ K$ when Sr is present. The radiative $V_{oc}$ limit for this cell is $V_{oc,rad} = \frac{k_B T}{e} ln\left(\frac{J_G}{J_{rad,0}}\right) = 1.337 eV$, where the dark radiative current $J_{rad,0}$ was calculated from the convolution of the $EQE_{PV}$ spectrum and blackbody radiation (Eq. S2, SI) at $300\ K$ (Fig. S3, SI), and $J_G$ is set equal to $J_{sc}$. Combining these values results in a predicted $V_{oc} = V_{oc,rad} - \Delta V_{oc,non-rad} = 1.337 - 0.161 = 1.176\ V$, which is very close to the measured averaged $V_{oc}$ for 2% Sr cells. The perovskite without Sr shows higher non-radiative voltage loss $\Delta V_{oc,non-rad} = 0.232\ eV$ with an almost identical radiative limit $V_{oc,rad} = 1.338\ eV$, leading to a predicted $V_{oc} = 1.338 - 0.232 = 1.106\ V$, again in very good agreement with the measured averaged $V_{oc}$ for 0% Sr cells. The non-radiative voltage losses $\Delta V_{oc,non-rad}$ of the two samples differ by 70 mV, matching exactly the $V_{oc}$ enhancement measured by *J-V* scans. The small difference of only 15meV found between the QFLS and the $V_{oc}$ for the Sr containing sample indicates that the energy losses due to the implementation of the charge transporting layer are successfully minimized.

**Morphology Characterisation**

As noted above, previous work showed that Sr segregates at the perovskite surface[22,23]. Fig. 6 shows the elemental distributions of several elements of our samples on ITO/glass as measured by secondary ion mass spectrometry (SIMS), utilizing O as primary ion source. According to Fig. 6 a) the elements comprising the perovskite, such as Cs, Pb, and Rb, are homogeneously distributed when Sr is not present. On the other hand, the SIMS in Fig. 6 b) clearly proves a significant enrichment of Sr at the surface and interface to ITO, with the Sr concentration being considerably lower in the bulk. In both measurements the signal of the negatively charged ions is low due to limitations of the sputtering yield; however those species have been detected with Cs ions as primary ion source and they have also shown homogeneous distribution (Fig. S4d, SI). To exclude the possibility of lateral inhomogeneities, we analysed different spots of ca. 250x250 µm$^2$ on the surface and from samples of different batches. In all measurements (Fig. S4a, S4b, S4c, SI), Sr traces show the same inhomogeneous distribution with a significant enrichment at the surface/interface. Following this picture, when the Sr is added to the precursor solution and the film is formed after spin coating, Sr clearly segregates the surface/interface, whereas the bulk remains almost unaltered. Additional XPS measurements (Fig. S5A, SI) show Sr traces on the sample surface with the rise of the characteristic Sr 3p peaks, in accordance to the SIMS results. A more detailed XPS surface composition analysis (Fig. S5B, SI) quantitatively shows that the relative atomic Sr concentration compared Pb is 10 times higher than what should be predicted by stoichiometry, in good agreement with the findings by Perez Del Rey et al. [22]. Detailed analysis of the core-level spectra results in the Sr/Pb molar ratio of 0.21 and the Sr/I molar ratio of 0.07, which are both much higher than the expected stoichiometry. This finding is in good agreement with SIMS traces and it confirms a Sr enriched region at the surface. Moreover, we can exclude a coverage of the surface with non-reacted SrI$_2$, since the I concentration found at the surface, and consequently its ratio with Sr, should be consistently higher than what found here. In addition, the I/Pb molar ratio, for the sample with Sr, is higher than sample without Sr, which can be possibly ascribed to the fact that Sr can partially replace Pb, leading to a decrease of Pb/I ratio at the surface in the perovskite lattice compared to what predicted by stoichiometry by a full Pb perovskite.

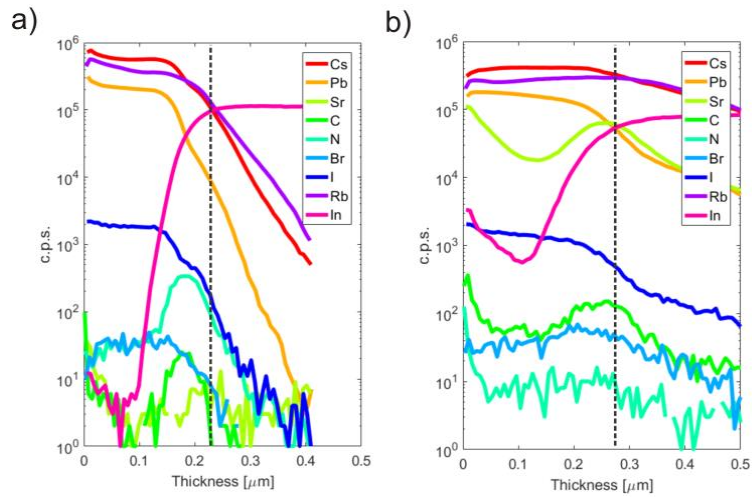

*Figure 6: Secondary ion mass spectrometry (SIMS) profiles for a specified group of elements as a function of depth for a neat perovskite layer on ITO/glass substrates a) with no Sr added and b) with 2% Sr. Note that in both figures a thickness equal to zero represent the top surface of the perovskite layer, while the rise of the indium signal (highlighted with a black dotted line) indicates that the perovskite layer is almost completely sputtered off and that the ITO substrate becomes exposed. The initial Indium signal for the Sr containing Sr might originate from pin holes or non-uniform film areas substrate and therefore being detected at the beginning of the scan. This agrees with the smeared profile of the signals in the Sr containing sample as an indication of a rougher surface as confirmed in SEM images in Fig. 8d.*

Scanning electron microscopy (SEM) confirms a notable change of the surface when Sr is added to the perovskite, as represented in Fig. S6A, SI. Upon Sr addition the surface starts to be characterized by leave-like bright areas. These brighter islands appear to be characterized by different work function after imaging them through energy sensitive SEM in-lens detector. We notice that these features can be resolved exclusively through this in-lens detector and not with an Everhart-Thornley lateral detector, more topographical sensitive. This excludes the possibility of having a non-conductive material covering the surface, otherwise appearing evidently with area of different brightness due to strong charging effect in both detectors. A comparison between the two different imaging techniques on the same surface spot is presented in Fig. S6B, SI. Features of similar size and shape have been reported previously on the surface of Sr-containing $CsPbI_2Br$ perovskite processed from solution[23]. The cross section (Fig. S6A d), SI) shows that grains propagate from the top to the bottom even in the presence of the additional features sitting on top of the layer. The results indicate that the top surface of the Sr-containing perovskite contains a material of a different composition spread across the surface area increasing the roughness. A series of images representing the surface after addition of Sr at different concentrations (Fig. S6C, SI) show that the density of these islands on the surface is directly correlated with the concentration of Sr added.

**Electronic Structure Characterisation**

The results above provide evidence that adding Sr to a quadruple cation perovskite has a significant effect on the perovskite surface, whereas the bulk properties seem to remain fairly unaltered, asking for a detailed investigation of the electronic structure of samples with and without Sr. Photoemission and inverse photoemission spectroscopy (PES and IPES) experiments were performed for solar cell related multilayer stacks comprising ITO, the hole-transporting PTAA, and the active perovskite, to retrieve information on surface work function (WF), ionization energy (IE), electron affinity (EA), as well as position of the valence band maximum (VBM) and the conduction band minimum (CBM) with respect to the Fermi level ($E_F$).

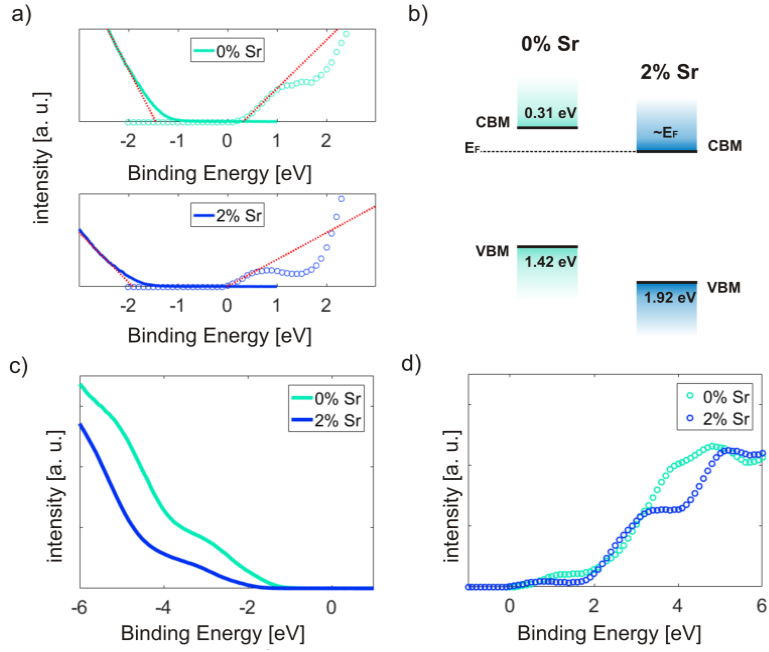

*Figure 7: UPS and IPES spectra for 0% Sr and 2% Sr containing perovskites. a) Magnified valence and conduction band regions near $E_F$. The binding energy scale is referenced to $E_F$, set to zero. b) Energy levels with respect to $E_F$ for 0% and 2% Sr perovskites obtained from UPS and IPES spectra. From this picture, we observe a strong n-type character of the perovskite surface, as well as a larger band gap for the case of 2% Sr. Wide range (c) valence and (d) conduction band regions.*

As shown in Fig. 7 a) and c), the valence band and VBM, when Sr is incorporated, are shifted to higher binding energy while the conduction band and CBM move closer to $E_F$ [Fig. 7 a) and d)]. VBM and CBM were evaluated by linear extrapolation of the valence and the conduction band onsets towards the background on a linear intensity scale, respectively, as shown in Fig. 7 a). For the 0% Sr-perovskite we extrapolated the VBM to be at 1.42 eV and the CBM at 0.31 eV relative to the Fermi level, giving a band gap of 1.73 eV. This value is comparable with the optical band gap of 1.63 eV reported for this type of quadruple cation perovskite[27]. We note that an overestimation of the band gap (from UPS and IPES) compared to the optical gap can be due to the linear extrapolation [42,43], and our results are consistent with examples reported in literature [42–45]. A lower band gap would be expected if the extrapolation was based on logarithmic plots, but due to the uncertainty of cross-section effects and rather large experimental broadening in IPES we refrain from such procedures here. Most importantly, we observe a strong effect on the electronic structure when Sr is incorporated. In particular, the CBM for 2% Sr perovskite locates at the Fermi level, indicating a strongly *n*-type surface. Concomitantly, the VBM is shifted towards higher binding energy, i.e., 1.92 eV (relative to $E_F$). From these measurements, we deduce that the Sr-enriched perovskite surface features a ca. 190 meV wider band gap compared to the surface of Sr-free perovskite, which is consistent with the smaller electronegativity of Sr compared to Pb. We notice that an increased band gap was predicted by DFT calculations for $Pb^{2+}$ being completely replaced by $Sr^{2+}$ in $MAPbI_3$ [21]. Since our samples have a different composition, a quantitative comparison of our results and theory is precluded.

The observation of a strongly n-doped surface of the Sr-containing sample raises the question whether the addition of Sr introduces doping also in the perovskite bulk. Unfortunately, ultraviolet (UV) photoemission is not suited to address this issue due to the short electron escape depth. However, given the small quantities of Sr added and strong segregation at the surfaces, bulk doping seems quite unlikely. In agreement with this interpretation is the pronounced concurrent increase in PL lifetime and absolute PL yield upon Sr addition as reported above, which excludes the presence of a high density of doping-induced background charges in the Sr-containing sample, similar to what has been reported by Bolink and coworkers on Sr-doped $MAPbI_3$ [22]. In order to support this claim, we simulated the TRPL decay and PL efficiency of the neat perovskite for varying doping concentration, using realistic parameters for the recombination coefficients (Fig. S7A, SI). According to these simulations, if the increase in PL efficiency is due to doping, this would cause a concurrent reduction of the PL lifetime, confirming previous studies[46,47]. In our case, the addition of Sr goes along with a significant increase of both PL decay times and PL efficiencies, ruling out extensive bulk doping in the presence of Sr. In fact, an effective lifetime of nearly 1 μs combined with a large QFLS suggests a doping density of less than $10^{15}$ $cm^3$, meaning that the perovskite bulk is nearly intrinsic.

The low doping density in the bulk combined with the strongly *n*-type surface suggests significant band bending to occur. To provide experimental evidence for this effect, surface photo voltage (SPV) measurements were performed on neat perovskite layers with and without Sr being present. Essentially, this technique determines the shift in work function (or VBM with respect to $E_F$) between dark and light measurements through Kelvin Probe or UPS. This experiment constitutes a common methodology to measure the degree of band bending in the near surface regions of semiconductors[48–50]. Moreover, it has been implemented in various studies on halide perovskites to quantify the degree of surface band bending[51–53]. Given the *n*-type nature of the surface studied here, it reasonable to assume that a considerable amount of donor like surface states donate electrons to the region close to the surface, leaving behind an accumulation of immobile positive charges at the surface. This gives rise to a surface space charge region with a downward band bending from the bulk towards the surface. Under illumination, photogenerated carriers are created due to band-to-band (or trap-to-band) transition. These new carriers will redistribute under the influence of the space charge field, thereby compensating the local excess of positive charges at the surface. As a result, illumination will gradually decrease the level of bend banding until flat band condition is reached, as schematically represented in Fig. 8a). Recently, surface band bending in perovskite has been proved with the same methodology used in our study for bare $MAPbI_xCl_{3-x}$ films. Here the effects of an *n*-type surface with downward band bending was clearly demonstrated[54] and attributed to the presence of donor levels at the perovskite surface, likely consisting of reduced lead ($Pb^0$).

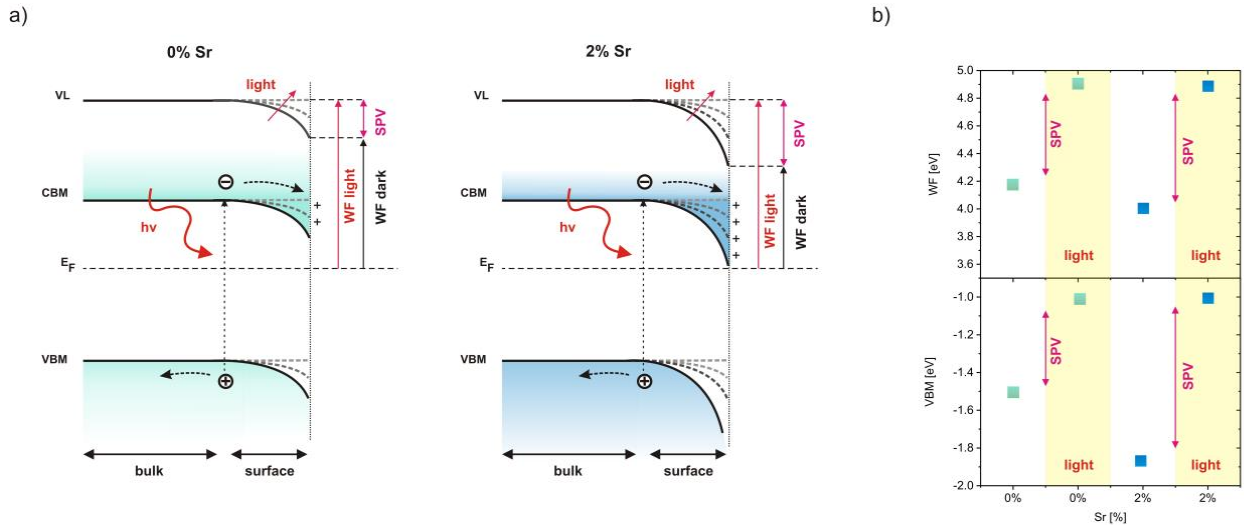

*Fig. 8: a) Schematic representation of the SPV effect of 0% and 2% Sr samples. Dotted lines represent the effect of white light illumination on the vacuum level (VL), conduction band minimum (CBM) and valence band maximum (VBM) in the surface-near region. In the limit of intense illumination, the space charge region is completely screened by photogenerated charge and flat band condition is established. The bulk and surface energetic in dark is represented by solid lines. In the diagram the SPV effects is highlighted only for WF shift, even though in the experiment also the VBM shift has been measured. Surface energy levels are from measurements of the work function and the valance band onsets in the dark, while bulk values are taken from the corresponding measurements under illumination, assuming that flat-band conditions are established as reported in Zu et al. [54]. b) Corresponding WF and VBM values from UPS, measured without and with simultaneous white light illumination for 0% and 2% Sr samples. Both samples display very similar VL and VBM positions under illumination, indicating nearly identical bulk energy levels for 0% and 2% Sr, while surface band bending is more significant in the Sr-containing sample, clearly observable due to a much larger SPV effect indicated with pink arrows.*

The effect of white light illumination on the WF and VBM is summarized in Fig 8 b) (see Fig. S8 SI for the corresponding UPS spectra). In the case of 0% Sr, the WF increases from 4.17 eV in dark to 4.93 eV under illumination, accompanied with the VBM shifting from 1.51 eV to 1.00 eV binding energy. VBM and ϕ do not shift perfectly in parallel, most likely due to surface inhomogeneities, which are differently affected by SPV, but cannot be differentiated in the area-averaging photoemission experiment. A much stronger SPV effect was observed for 2% Sr samples where WF shifted from 4.00 eV in dark to 4.89 eV under illumination, and the VBM from 1.87 eV to 1.03 eV, respectively. We note that the SPV was reversible for multiple illumination/dark cycles, thus photochemical reactions and degradation can be excluded. Additionally, in this measurements we notice a reduction in work function after Sr addition, in agreement with the effects previously reported in literature[22]. A change of sample work function can have manifold reasons, e.g., due to formation of dipoles at the surface[42], a change in stoichiometry[55,56], or it can be indeed associated with a more *n*-doped surface[57]. In conclusion, UPS reveals a much larger degree of surface band bending of the Sr containing sample, resulting in a stronger SPV effect compared to 0% Sr sample, see magenta coloured arrows. At the same time, both samples exhibit very similar values of the WF and VBM under illumination, meaning that the bulk energetic is almost independent of the Sr content. This indicates that the doping caused by Sr addition is limited to the surface, with very little bulk doping, in full accordance with the results from transient and steady state PL as reported above.

**Discussion**

Before proposing a model to explain the beneficial effect of adding Sr to a quadruple cation perovskite, we first summarize the key findings from our studies (Table 2). From SIMS and XPS, we find a strong enrichment in Sr at the surface denoted in a Sr/Pb ratio of 0.21 being much higher than the expected 0.02 from a homogeneous Sr addition assumption. In TRPL, we find that Sr addition strongly suppresses the initial fast PL decay, indicating reduced trapping, but also prolongs the long-term decay attributed to non-radiative recombination. In accordance with this, we observe a ca. 30 meV increase in the quasi-Fermi level splitting in the neat perovskite under 1 sun equivalent illumination conditions if 2 % Sr is added. In the complete device, Sr suppresses non-radiative recombination, strongly enhancing the electroluminescence efficiency 25-fold and reducing the non-radiative voltage loss from 230 to 160 meV. On the other hand, very similar absorption and EQE$_{PV}$ properties for all devices suggest only a small effect of Sr addition on the bulk properties, consistently with SPV measurements where upon illumination bands flatten at the same position but a much stronger surface workfunction shift is measured denoting stronger band bending with Sr. Finally these properties lead to the measured enhancement of 70 mV in $V_{oc}$.

*Table 2: Summary of the most relevant photovoltaic and optoelectronic properties determined for 0% Sr and 2% Sr containing perovskite films and/or solar cells. The insertion of Sr results in a consistent improvement of all parameters.*

| Sr [%] | Sr/Pb surface ratio | $\tau_{eff}$ [ns] | $\Delta E_F$ [eV] | EQE$_{EL}$ [%] | $\Delta V_{oc,non-rad}$ [eV] | WF shift under illum. [eV] | VBM shift under illum. [eV] | $V_{oc}$ [V] |
|---|---|---|---|---|---|---|---|---|
| 0 | 0 | 180 | 1.16 | 0.01 | 0.23 | 0.76 | 0.51 | 1.11 |
| 2 | 0.21 | 980 | 1.19 | 0.25 | 0.16 | 0.89 | 0.84 | 1.18 |

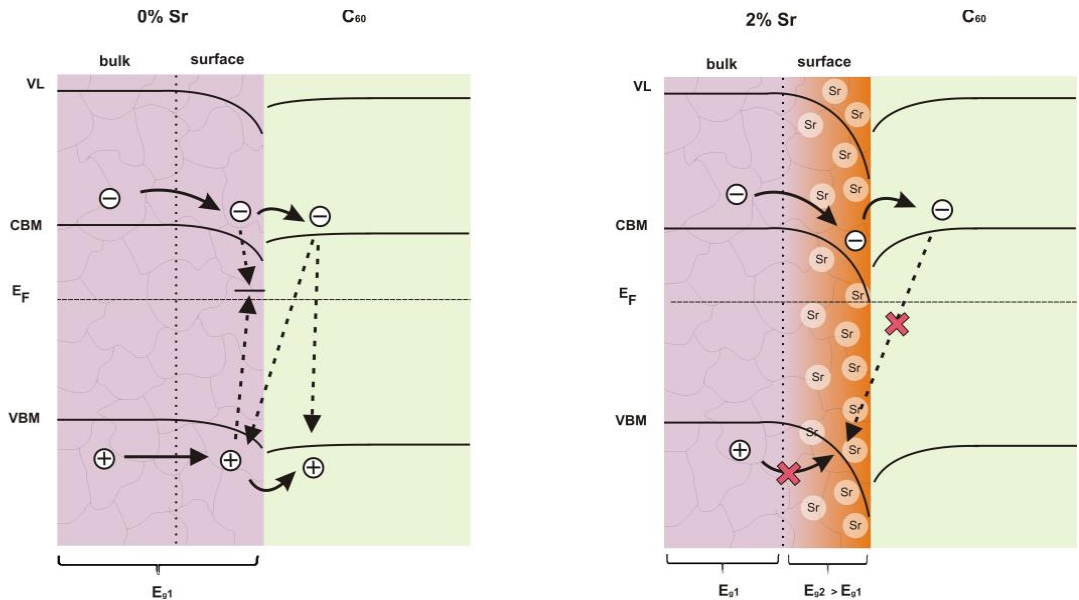

*Figure 9: Schematic representation of the two perovskite samples with and without Sr addition with respective recombination schemes. From a combination of morphological studies, PES and SPV experiments it is possible to draw a picture where Sr segregates at the perovskite surface inducing a wider band gap and a more*

*pronounced n-type character. On the other hand, optical measurements and SPV suggests that the bulk shows very similar characteristics for both samples. The energy levels in the bulk and at the surface for the 0% Sr and the 2% Sr cases, respectively, are based on values from UPS and IPES and taking into consideration the surface band bending found from SPV measurements. In the sample without Sr the recombination of charges can happen at the surface or across the interface with $C_{60}$. The Sr enriched surface repels holes from the surface reducing the non-radiative recombination at or across the interface and improve the selectivity of the contact with $C_{60}$. Energy level diagrams were drawn with UPS results on 20 nm $C_{60}$-covered perovskites, see Fig. S7a and S7b, SI. The trap state present in the 0% Sr sample is absent in the 2% Sr in order to indicate a possible passivation effect of surface islands.*

In Fig. 9 we schematically represent two perovskite samples, without and with Sr, combining findings from all previous studies as a summarizing figure. We propose that this specific energetic landscape is responsible for a considerable reduction of surface recombination as indicated in Fig. 9. Notably, a strong *n*-type character of the surface goes along with a nearly complete occupation of the electron traps already in the dark, reducing the probability that photo-generated electrons become trapped by these states, fully consistent with the almost complete absence of an initial fast PL decay in the Sr-containing layer. More important, the stronger surface band bending, as measured in SPV when Sr is added, repels holes from the surface, with the benefit of limiting the probability of a captured electron in a surface state to recombine with a hole. The most well-known example of this phenomena is the so-called "back-surface field"[58] which is an established approach for suppressing surface recombination in inorganic semiconductors. In addition, the accessibility of the surface for electrons and/or holes may be reduced by the wider band gap of the Sr-containing surface perovskite region, similar to what has been proposed recently for a compositionally engineered perovskite/HTL interface[15]. Additionally, the Sr induced surface modification and the formation of the islands on the top surface may also act as a passivating agent, reducing the number of interfacial states responsible for trap-assisted recombination. Even though these three effects cannot be disentangled based on the presented set of experiments, we can conclude that these are all possible beneficial effects due to the addition of Sr. Additionally, they all may act in combination together to reduce non-radiative recombination in actual devices by strongly suppressing interface recombination with $C_{60}$, as proposed in Fig. 9.

In order to prove the beneficial effect of Sr in reducing non-radiative recombination at the perovskite/$C_{60}$ surface, absolute PL efficiencies were measured for perovskite films on glass, without and with 2 % Sr, additionally covered with a 30 nm thick layer of $C_{60}$ (see Fig. S10, SI for the results). As expected, the measurements reveal a substantial decrease of the PL efficiency in the presence of $C_{60}$ for both samples. However, when Sr is present the decrease in PL attributed to the presence of $C_{60}$ is strongly reduced. This confirms our proposal that Sr addition improves the device performance by reducing the strength of the non-radiative recombination specifically at the surface of the absorber and at the interface with $C_{60}$. However, the modification of the interface morphology and energetics upon Sr addition might induce an extra barrier for extraction of electrons or holes, as extensively discussed for Fig. S9, SI, which may be a main cause for the systematic decrease in FF. This will be subject to follow-up work.

## **Conclusions**

In summary, we show that the addition of a small amount of $SrI_2$ to the precursor solution of a hybrid quadruple cation perovskite induces are remarkable increase of the $V_{oc}$ without the need of any post-deposition treatment or interlayer deposition. The attained $V_{oc}$ of 1.18V is among the highest of *pin* perovskite solar cells with a $C_{60}$ electron-transporting layer for the given bandgap of 1.62 eV. We note that during revision of this manuscript, a publication has appeared reporting a $V_{oc}$ of up to 1.21 V for a similar *pin* perovskite architecture, utilizing a 2-step composition engineering to alter the energetics of the perovskite surface[59], not required here. Transient and absolute PL measurements in combination with PES suggests that the addition of Sr causes a drastic reduction of non-radiative recombination at the interface between the perovskite and the $C_{60}$ electron-transporting layer. We assign this improvement to a preferential *n*-doping of the perovskite surface and to the concurrent formation of a space-charge region which, combined with a locally larger band gap at the surface, reduces the access of photogenerated holes to the perovskite surfaces, similarly to the well-known back-field effect in inorganic semiconductor. Adding a thin layer of an insulating polymer on top of the Sr-enriched perovskite surface improved the $V_{oc}$ further to nearly 1.23V, corresponding to a non-radiative voltage loss of only 110 meV. With that, our work not only demonstrates very high $V_{oc}$ values and efficiencies in Sr-containing quadruple cation perovskite *pin* devices, but also highlights the importance of addressing and minimizing the losses located at the interface with the transport layers in working solar cells.

## Materials and Methods

### Device Preparation

Patterned indium-doped tin oxide (ITO, Lumtec, 15 ohm sqr.$^{-1}$) was washed with acetone, Hellmanex III, DI-water and isopropanol. After microwave plasma treatment (3 min at 200 W) poly[bis(4-phenyl)(2,4,6-trimethylphenyl)amine] (PTAA, Sigma-Aldrich, Mn = 7000–10000, PDI = 2–2.2) in a concentration of 1,5 mg/ml was spin-coated at 6000 rpm for 30 seconds and immediately annealed for 10 minutes at 100º C. The perovskite layer was formed by spin coating a DMF:DMSO solution (4 :1 volume) at 4500 rpm for 35 seconds. After 10 seconds of spin coating, 500 mL of diethyl ether (antisolvent) was dripped on top of the spinning substrate. After spin coating samples were annealed at 100º C for 1 h. Afterwards, samples were transferred to an evaporation chamber and $C_{60}$ (30 nm), BCP (8 nm) and copper (100 nm) were deposited under vacuum ( p =$10^{-7}$ mbar). The active area was 6 mm$^2$ defined as the overlap of ITO and the top electrode.

**Current Density–Voltage Characteristics and EQE$_{PV}$**

J-V curves were measured under N$_2$ on a Keithley 2400 system in a 2-wire configuration with a scan speed of 0.1V/s and voltage step of 0.02V. One sun illumination at approximately 100mWcm$^{-2}$ of AM1.5G irradiation was provided by a Oriel class ABA sun simulator. The real illumination intensity was monitored during the measurement using a Si photodiode and the exact illumination intensity was used for efficiency calculations. The sun simulator was calibrated with a KG5 filtered silicon solar cell (certified by Fraunhofer ISE). The AM1.5G short-circuit current of devices matched the integrated product of the EQE spectrum within 5-10% error. The latter was recorded using a home build set-up utilizing a Philips Projection Lamp (Type7724 12 V 100 W) in front of a monochromator (Oriel Cornerstone 74100) and the light was mechanically chopped at 70 Hz. The photo-generated current was measured using a lock-in-amplifier (EG&G Princeton Applied Research Model 5302, integration times 300 ms) and evaluated after calibrating the lamp spectrum with an UV-enhanced Si photodetector (calibrated at Newport).

**Photoluminescence and Electroluminescence**

Timeresolved PL data was acquired with a TCSPC system (Berger & Lahr) after excitation with a pulse-picked and frequency-doubled output from a mode-locked Ti:sapphire oscillator (Coherent Chameleon) with nominal pulse durations ~ 100fs and fluence of ~30nJ/cm² at a wavelength of 470nm.

Hyperspectral Absolute Photoluminescence Imaging was performed by excitation with two 450 nm LEDs equipped with diffuser lenses. The intensity of the LEDs was adjusted to ~1 sun by illuminating a contacted perovskite solar cell (short circuit) and matching the current density to the short circuit current measured in the *JV* sun simulator (The measured short circuit current density of the solar cell under this illumination was 22.2 mA/cm$^2$. The photoluminescence image detection was performed with a CCD camera (Allied Vision) coupled with a liquid crystal tuneable filter. The system was calibrated to absolute photon numbers in two steps in a similar way to the process described by Delemare et al.[60] For this purpose an IR laser diode and a spectrally calibrated halogen lamp was coupled to an integrating sphere. The pixel resolution of the images corresponds to about 10 µm in diameter. Sets of images from 650 nm to 1100 nm with 5 nm step size were recorded. All absolute PL measurements were performed on films with the same thicknesses as used in the operational solar cells.

The EL spectra were acquired using an Andor SR393i-B spectrometer equipped with a silicon detector DU420A-BR-DD (iDus). The response of this setup was measured with a calibrated lamp (Oriel 63355). The cells were kept under forward bias with a Keithley 2400 at an applied current $J_{inj} \approx J_{sc}$, stabilized for 20 seconds before recording a spectrum. Absolute EL was measured with a calibrated Si photodetector (Newport) attached to a Keithley 485 pAmeter. The photodetector was placed directly in front of the device pixel, then a forward bias was applied with a Keithley 2400 source-meter and the resulting photonflux was calculated considering the EL spectrum of the solar cell and the spectral response of the Si photodiode. Injected current and photodector response ware monitored with a home written LebVIEW routine varying the voltage and stabilising for 20 seconds after every voltage step (dV typically 0.02V).

**Morphological Characterization**

SIMS measurements were performed using a Cameca IMS4f instrument, using $O^{2+}$ and $Cs^+$ as primary sputtering ions within an energy range of 5-15 keV. The scanned area on the sample surface was 250x250 $\mu m^2$.

SEM images were acquired with a Zeiss Ultra Plus SEM. Images has been acquired through the use of Secondary Electron in-lens detector.

**Photoemission and Inverse Photoemission Spectroscopy Measurements:**

Photoemission experiments were performed at an UHV system consisting of sample preparation and analysis chambers (both at base pressure of $1 \times 10^{-10}$ mbar), as well as a load-lock (base pressure of $1 \times 10^{-6}$ mbar). All the samples were transferred to the UHV chamber using a transfer rod under rough vacuum ($1 \times 10^{-3}$ mbar). UPS was performed using helium discharge lamp (21.22 eV) with a filter to reduce the photo flux and to block visible light from the source to hit the sample. XPS was performed using Al K*α* radiation (1486.7 eV) generated from a twin anode X-ray source. All spectra were recorded at room temperature and normal emission using a hemispherical SPECSPhoibos 100 analyzer. All perovskite layers were deposited on a ITO / PTAA stack in order to be as much close as possible to real device conditions. The resolution and energy calibration of the PES and IPES were determined by measuring the Fermi edge of a clean Au (111) single crystal. The overall energy resolution was 140 meV and 1.2 eV for UPS and XPS, respectively. The IPES measurements were performed in the isochromat mode using a low-energy electron gun with a BaO cathode and a band pass filter of 9.5 eV ($SrF_2$ + NaCl). All presented PES and IPES spectra are given in binding energy (BE) referenced to the Fermi level. The overall energy resolution for IPES was 0.74 eV. The UPS spectra of thin films under consecutive on-off illumination circles were conducted using a white halogen lamp at 150 mW/cm$^{-2}$ (daylight rendering spectrum). The same experimental conditions and setup presented in the work of Zu et al. [54] has been used for our set of measurements. C60 molecules were purchased from *Novaled*, and were used as received and thermally evaporated from resistively heated quartz crucibles. The nominal deposited thickness was monitored by a quartz crystal microbalance.

**Competing financial interests**

The authors declare no competing financial interests.

**Acknowledgements**

S.A. acknowledges funding from the German Federal Ministry of Education and Research (BMBF), within the project "Materialforschung für die Energiewende" (grant no. 03SF0540), and the German Federal Ministry for Economic Affairs and Energy Energy (BMWi) through the "PersiST" project (grant no. 0324037C). Additional funding came from HyPerCells (a joint graduate school of the Potsdam University and the HZB) and the German Research Foundation (DFG) within the collaborative research center 951 "Hybrid Inorganic/Organic Systems for Opto-Electronics (HIOS)". We thank RTG Mikroanalyse GmbH, Schwarzschildstraße 1, 12489 Berlin for performing the SIMS measurements. The authors thank Sebastián Caicedo-Dávila and Dr. Ulrich Hörmann for fruitful discussions.

# High Open Circuit Voltages in *pin*-Type Perovskite Solar Cells through Strontium Addition

Pietro Caprioglio [⁂†], Fengshuo Zu [ǂ‡], Christian M. Wolff [⁂], José A. Márquez Prieto [ǂ], Martin Stolterfoht [⁂], Norbert Koch [ǂ‡], Thomas Unold [ǂ], Bernd Rech [§], Steve Albrecht [†§] and Dieter Neher [‡]

⁂ University of Potsdam, Institut für Physik und Astronomie, Potsdam, Germany
‡ Humboldt-Universität, Institut für Physik, Berlin, Germany
§ Helmholtz-Zentrum Berlin, Institute for Silicon Photovoltaics, Berlin, Germany
†Helmholtz-Zentrum Berlin, Young Investigator Group Perovskite Tandem Solar Cells, Berlin, Germany
ǂHelmholtz-Zentrum Berlin für Materialien und Energie GmbH, Berlin, Germany

**Supporting Information**

1) Current density – voltage characteristic
2) EQE and Absorption Measurement
3) Radiative Losses
4) SIMS (Secondary Ion Mass Spectroscopy)
5) XPS (X-Ray Photoelectron Spectroscopy)
6) SEM (Scanning Electron Microscopy)
7) Recombination Dynamic Simulations
8) Surface Photovoltage Effect
9) Photoemission Spectroscopy
10) Photoluminescence Quantum Yield

# 1. Current density – voltage characteristic

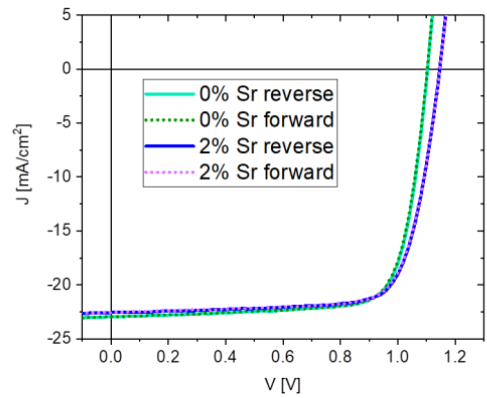

*Figure S1A*: *J-V characteristic showing both forward and reverse scan at 0.1V/s with a voltage step of 0.02 V for two samples containing 0% ans 2% Sr respectively. Both samples show complete absence of any hysteresis effect.*

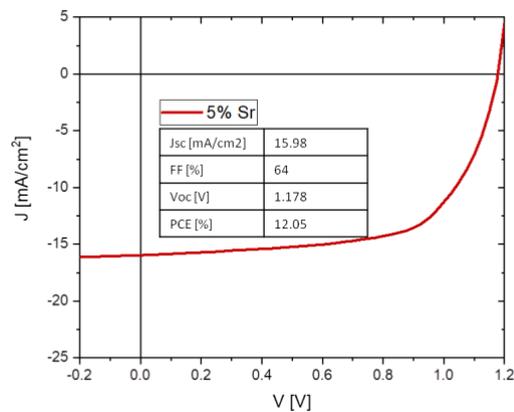

*Figure S1B*: *J-V characteristic showing a reverse scan at 0.1V/s with a voltage step of 0.02 V for a sample containing 5% Sr. The curve shows how a higher Sr concentration has negative effects on both $J_{sc}$ and FF without any appreciable improvement of the $V_{oc}$.*

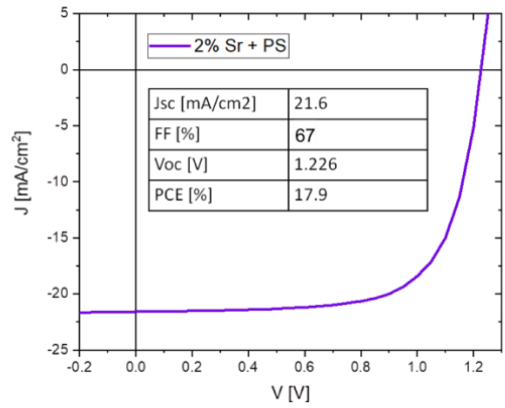

*Figure S1C:* J-V characteristic showing a reverse scan at 0.1V/s with a voltage step of 0.02 V for a sample containing 2% Sr with additional ultra-thin (less then 5nm) polystyrene layer deposited between perovskite and $C_{60}$ following the experimental procedure described in Wolff et al.[1] The resulting solar cell show an extraordinarily high Voc of 1.215V. Unfortunately, the presence of this insulating PS layer led to a considerable reduction of the FF, probably by limiting the extraction of charges via tunneling[1]

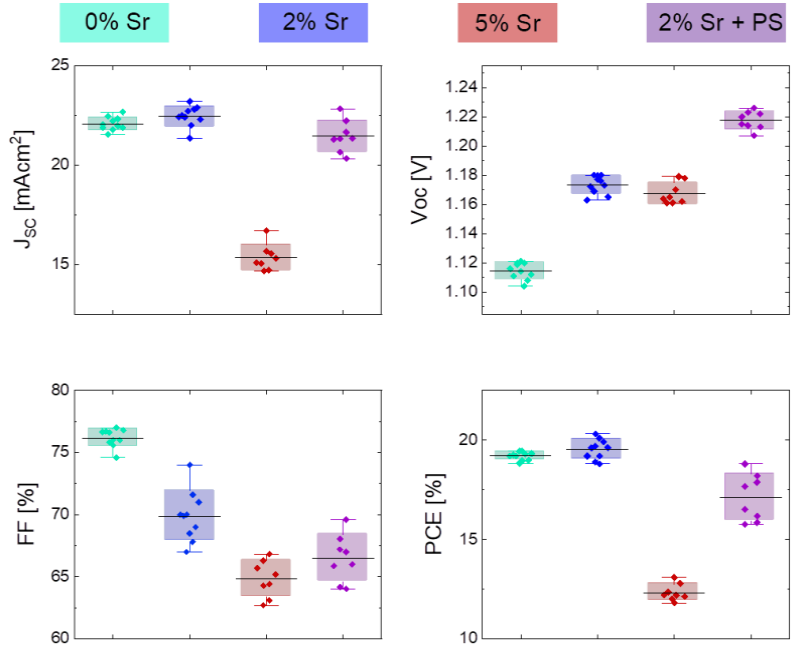

*Figure S1D:* Box chart for the most relevant type of devices, 0% Sr, 2% Sr, 5% Sr and 2%Sr + PS. From the picture trends are more evident: it is clear that upon Sr addition the $V_{oc}$ is improved whereas the FF decreases. We can identify a ratio of 2% Sr as the optimum concentration considering the strong decrease in

*performances observed with a 5% Sr ratio, without showing any further improvement in $V_{oc}$. On the contrary, the addition of a polystyrene layer onto a 2% Sr increase considerably further the Voc reaching a record value of 1.226 V. However, it is also clear of this has a negative effect on the FF and the overall PCE unfortunately is not improved. For this reason we refer to this additional increase in $V_{oc}$ as a proof of concept for the suppression of recombination at the interface with $C_{60}$.*

## 2. EQE and Absorption Measurement

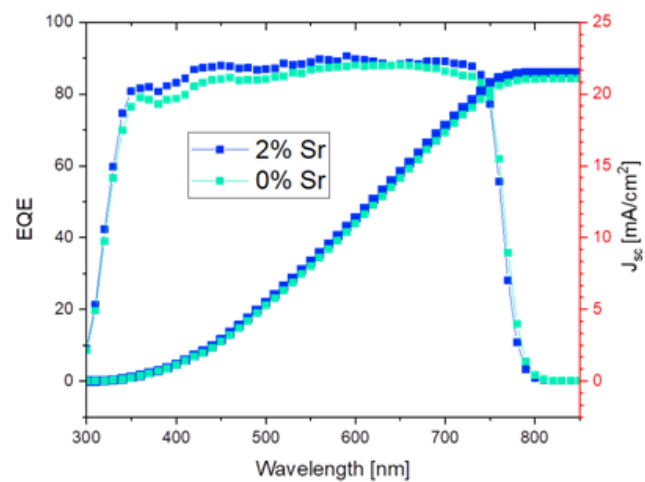

*Figure S1A: External Quantum Efficiency including the integrated current for 0% and 2% Sr cells. We acknowledge that for the 2% Sr cell the integrated current of 21.5 mA/cm² represents approximately a 4% relative mismatch compared to the $J_{sc}$ of 22.4 mA/cm² as obtained from the JV scan of the corresponding solar cell. A similar mismatch has been systematically found for the 0% Sr cell.*

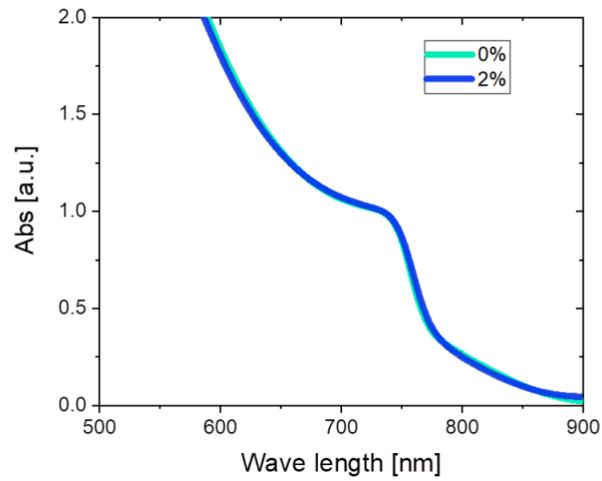

*Figure S2B*: Absorption measurement of two perovskite film with same thickness (400nm) with 0% and 2% Sr respectively, deposited on glass.

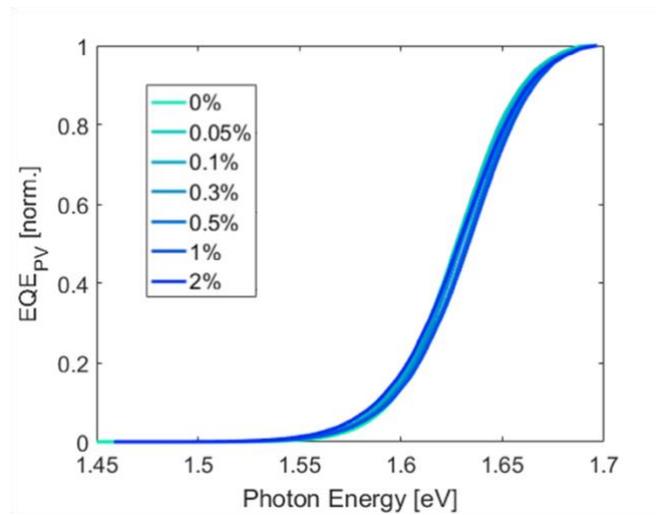

*Figure S2C*: $EQE_{PV}$ onset as function of photon energy for all different Sr concentrations, normalized to 1.7 eV.

### 3. Radiative losses

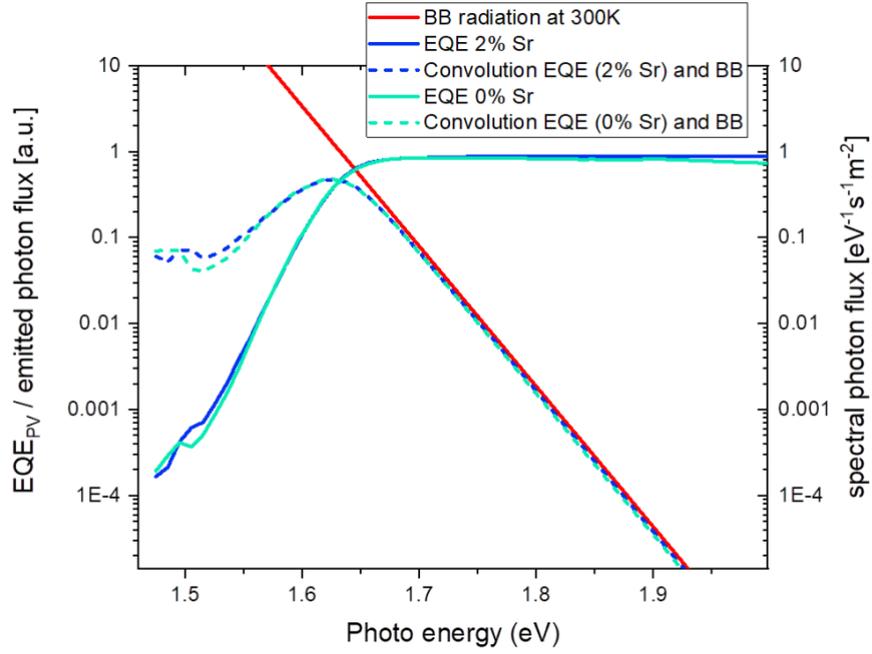

*Figure S3: EQE$_{PV}$ onset of two perovskite solar cell containing 0% and 2% Sr respectively and their emitted spectral photon flux calculated when the device is in equilibrium with the black-body (BB) radiation of the surroundings at 300K according to equation S1.*

Approach used in Fig. S6 follows the report by Rau et al. [2]. Briefly the Black Body photon flux is

$$\phi_{BB} = \frac{1}{4\pi^2 \hbar^3 c^2} \frac{E^2}{\exp\left(\frac{E}{k_B T}\right) - 1} \qquad (S1)$$

with $\hbar$ Plank's constant, $k_B$ Boltzmann constant and T temperature. Assuming that the perovskite solar cell is at 300K in thermal equilibrium with its environment, the dark radiative recombination current is:

$$J_{em,0} = e \int EQE_{PV}(E) \phi_{BB}(E) dE = J_{rad,0} \qquad (S2)$$

with $EQE_{PV}$ the photovoltaic external quantum efficiency of the perovskite solar cell and $J_{em,0}$ the current giving rise to emission, which also defines the dark radiative recombination current at $V = 0$. From that, the radiative $V_{oc,rad}$ ($EQE_{EL} = 1$) can be calculated with the following equation:

$$V_{oc,rad} = \frac{k_B T}{e} \ln\left(\frac{J_G}{J_{rad,0}} + 1\right) \quad (S3)$$

where $J_G$ is the generation current under illumination, in this case approximated to the short circuit current $J_{sc}$. For a more detailed derivation refer to Ref. 2.

## 4. SIMS (Secondary Ion Mass Spectroscopy)

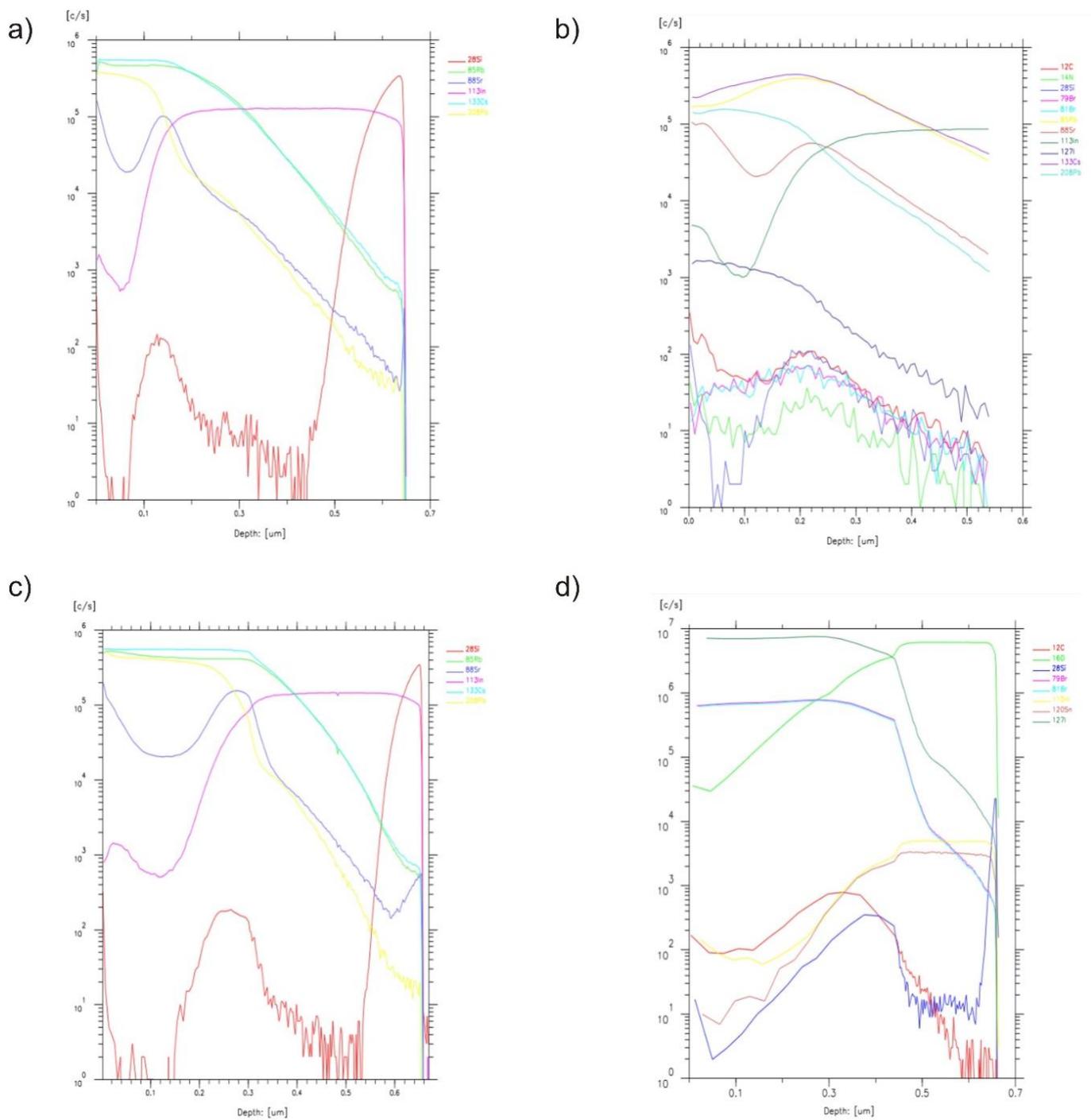

*Figure S4: a) and b) SIMS (Secondary Ion Mass Spectroscopy) depth profile of two different spots on the surface of the same perovskite sample containing 2% Sr. c) SIMS depth profile of the different perovskite sample containing 2% Sr. All measurements here are performed using O ions. d) SIMS depth profile using Cs ions in order to detect the distribution of Br and I across the perovskite layer.*

## 5. XPS (X-Ray Photoelectron Spectroscopy)

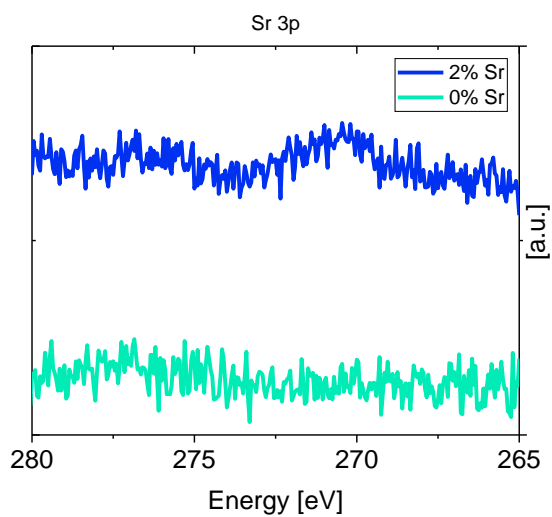

*Figure S5A*: XPS measurements of the surface of two perovskite layers containing 0% and 2% Sr respectively.

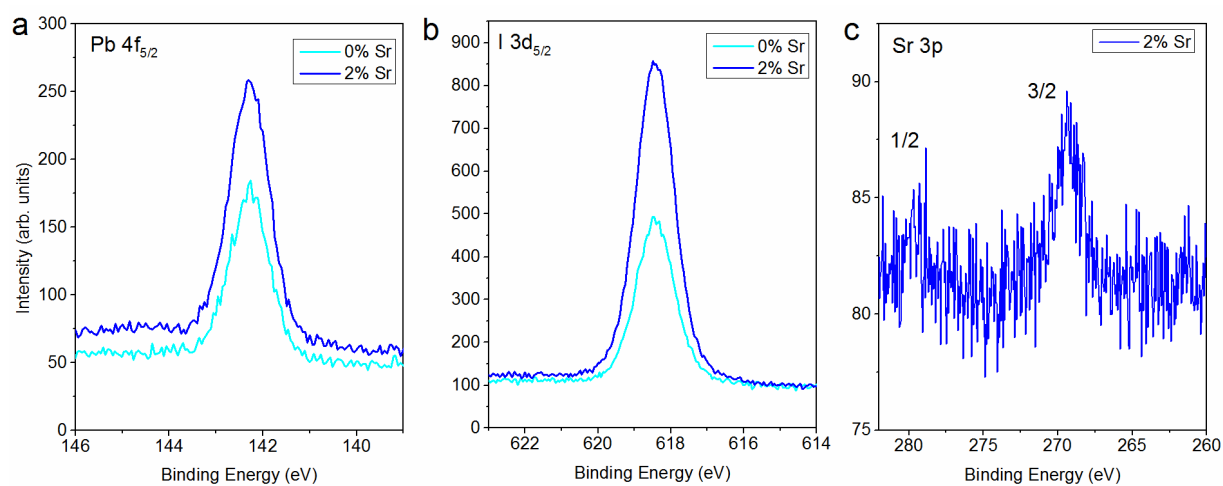

*Figure S5B*: XPS core-level spectra of (a) Pb 4f, (b) I 3d and (c) Sr 3p on two mixed perovskite films without and with 2% Sr incorporation.

In order to investigate quantitatively the surface composition of the perovskite films, XPS measurements employing monochromated Al kα radiation were performed on the samples without and with Sr incorporation. As shown in Figure S5, clearly, we observed relatively large amount of Sr on the perovskite surface with 2% Sr. Detailed analysis of the core-level spectra results in the Sr/Pb molar ratio of 0.21 and the Sr/I molar ratio of 0.07, which are both much higher than the expected stoichiometry from an homogenous Pb substitution through the whole volume. This proves strong enrichment of Sr on the perovskite surface, in good agreement with the findings by Perez Del Rey et al.[3]. Moreover, if the surface would be covered with unreacted $SrI_2$ the I concentration found at the surface, and consequently its ratio with Sr, should be consistently higher than what found here. In addition, the I/Pb molar ratio (2.49 by stoichiometry) is estimated to be 2.78 for the sample with Sr, in contrast to 2.13 for the sample without Sr, which can be possibly ascribed to the fact that Sr partially replace Pb, leading to a decrease of Pb/I ratio at the surface in the crystal lattice.

## 6. SEM (Scanning Electron Microscopy)

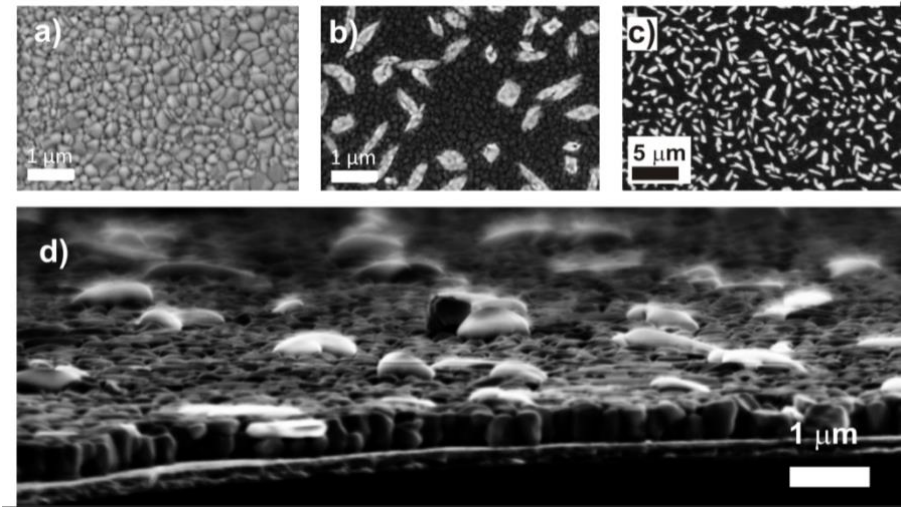

*Figure 6A:* Energy sensitive SEM in-lens detector top view images of perovskite films with a) 0% Sr and b) 2% Sr. c) shows a zoomed-out view of the same sample as in b). Image d) displays SEM in-lens detector images of a tilted sample containing 2% of Sr to visualize the cross section.

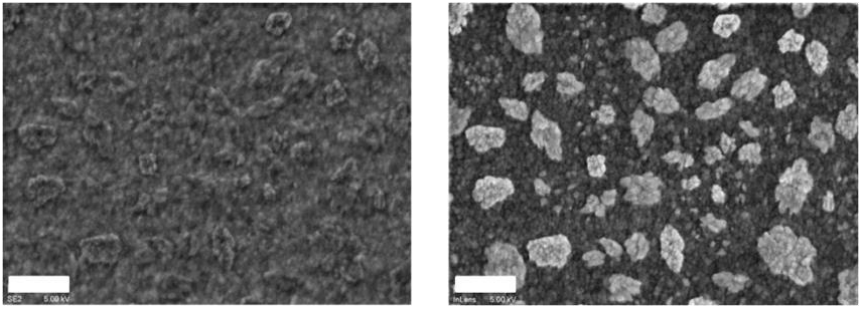

*Figure S6B*: Comparison between Everhart-Thornley detector (left) and In-Lens detector (right) imaging of a 2% Sr sample. The images show that the brighter feature on the top surface can be imaged only through the energy sensitive In-Lens detector whereas with the more topographic sensitive detectors those are barely visible. The comparison shows how from a more topography sensitive imaging it is possible to barely distinguish these features only by a light shadowing effect due to the different height compared to the rest of the layer, but no change in brightness is observed, excluding the presence of non-conductive material. Scale bar here is 1 $\mu$m.

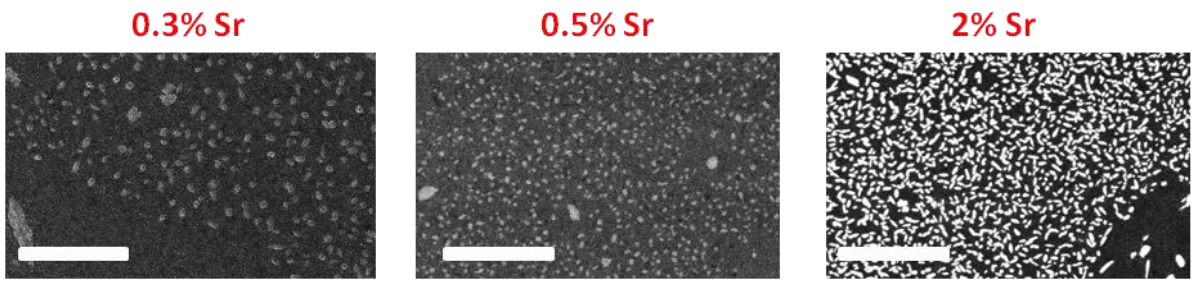

*Figure S6C:* Top SEM images of 0.3%, 0.5% and 2% Sr samples. The series of images indicates that increasing the Sr concentration the density of the brighter feature on top increase. Scale bar here is 10 $\mu$m.

## 7. Recombination Dynamic Simulations

In this simulation the recombination of charges has been simulated using the following rate equation

$$\frac{dn}{dt} = -(k_1 n + k_2(n_0 + n) \cdot n + k_3 n^3)$$

*(S4)*

where *n* is the initial photogenerated carrier density, $n_0$ is the background carrier concentration due to doping, and $k_1$, $k_2$ and $k_3$ are the monomolecular, bimolecular and Auger recombination coefficient, respectively. The simulations were carried out in MATLAB R2017b, running an iterative code with the given constants and equations. To obtain the steady state carrier concentration we assumed $G = 4 * 10^{21}$ cm$^{-3}$s$^{-1}$ and let the time run for ~10 ms until steady state conditions were assured.

Figure S7A displays the result of simulations where the background carrier concentration no was varied. For the pulsed simulation we assumed $n_{t=0} = 1 * 10^{14}$ cm$^{-3}$ at t = 0 and no further generation thereafter. In this case we simulated the recombination dynamic varying the doping concentration $n_0$ from $n_0 = 1 * 10^{14}$ cm$^{-3}$ to $n_0 = 1 * 10^{18}$ cm$^{-3}$, whereas the recombination constants fixed at $k_1 = 4 * 10^6 \, s^{-1}$, $k_2 = 1 * 10^{-10} \, s^{-1}$ and $k_3 = 1.8 * 10^{-28} \, s^{-1}$, being those realistic values and similarly already reported in literature[4,5]. The PL efficiencies at a carrier concentration equivalent to 1 sun has been simulated using

$$PL = k_2 \cdot (n + n_0) \cdot n$$

*(S5)*

and

$$PLQY = \frac{k_2 \cdot (n + n_0) \cdot n}{(k_1 n + k_2(n_0 + n) \cdot n + k_3 n^3)}$$

*(S6)*

The simulations show that increasing the background density due to doping provides additional centres for radiative recombination (we assumed exclusively radiative second order recombination), with the effect of increasing the radiative efficiency but at the same time enhancing the speed of the recombination of photogenerated charges, resulting in faster PL decays. Cleary, a very different situation is found in our study when increasing the Sr concentration, where we observe a parallel increase of the PL life times and PL efficiencies.

To explain the experimental results, a second set of simulations where performed where the PL decay rate and absolute PL efficiency was simulated for different $k_1$ in absence of doping. In this case simulation shows how the reduction of $k_1$ has positive effects on both PL decays and efficiencies.

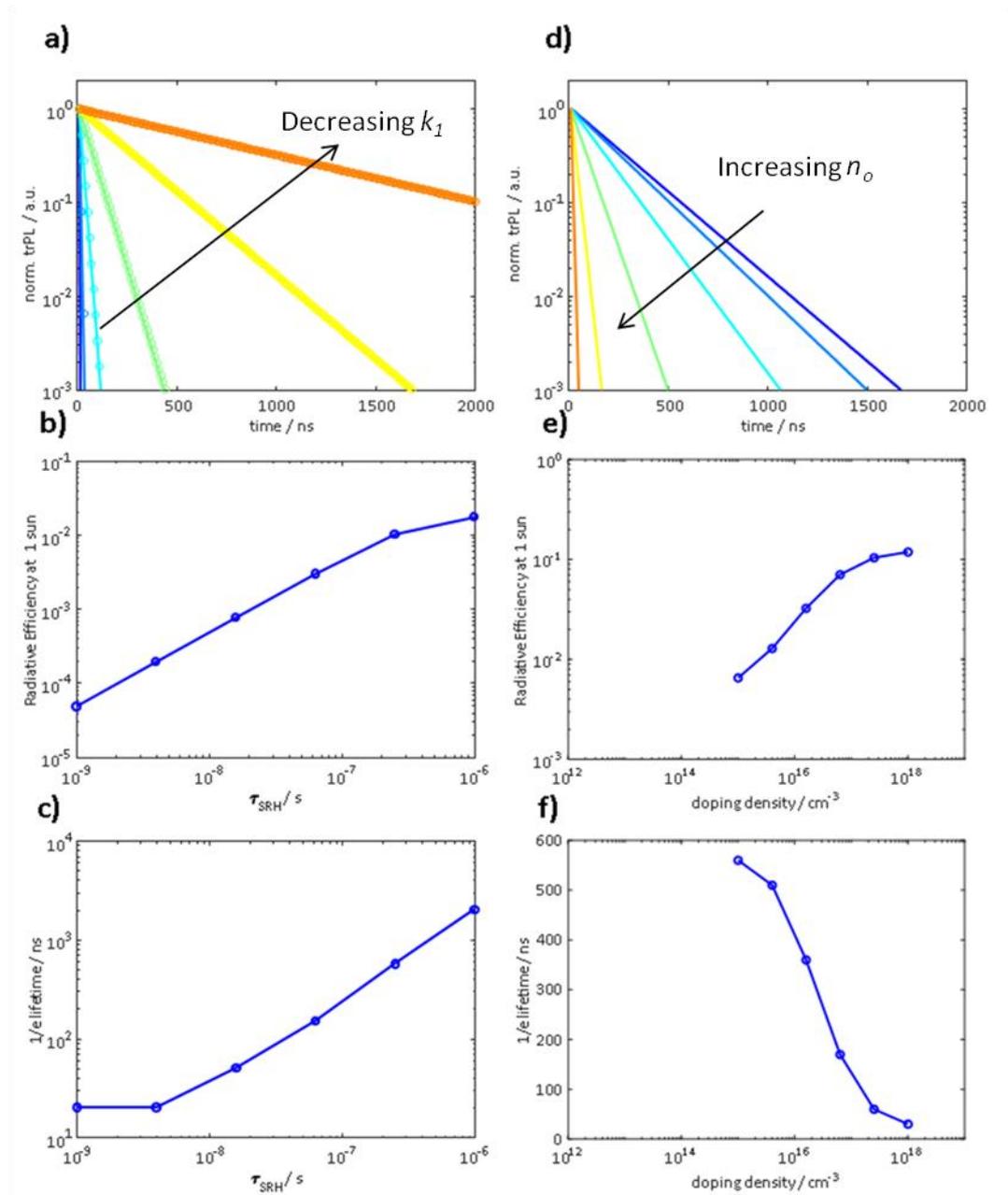

*Figure S7A:* Simulations of photoluminescence decays at low intensities ($n_{t=0} = 10^{14} cm^{-3}$), a) and d), radiative efficiencies (PLQY), b) and e), and 1/e PL decay, c) and f). On the left side, a), b) and c), the simulations present the results for reducing the monomolecular rate constant $k_1$ from $10^9$ and $10^6$ $s^{-1}$ and plotted against the SRH life times $\tau_{SRH} = 1/k_1$. On the right side, d), e) and f), we increase the background doping concentration from $10^{15}$ to $10^{18}$ $cm^{-3}$, using a fixed $k_1 = 4 * 10^6 \, s^{-1}$.

## 8. Surface Photovoltage Effect (measured by UPS)

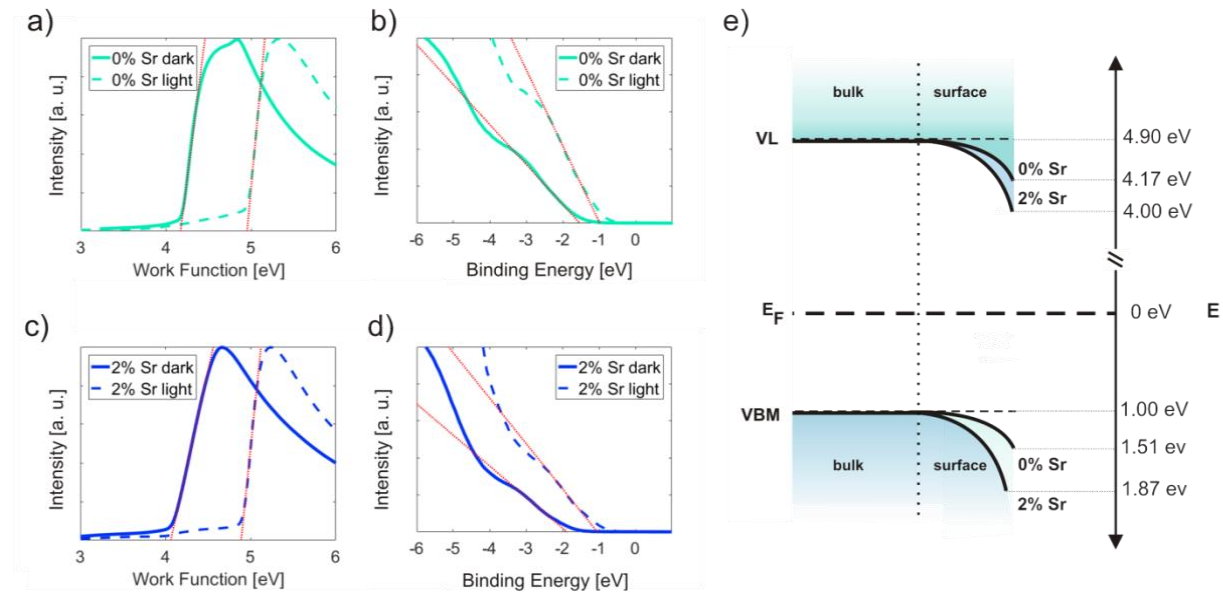

*Figure S8:* Effect of visible light illumination on the work function and valence band for 0% Sr and 2% Sr containing samples. a) and c) work function measured under dark and illumination conditions. b) and d) valence band regions measured under dark and illumination conditions. e) schematic representation of the surface band bending for 0% and 2% Sr-perovskites in the dark and under illumination. Vacuum level (VL) and valence band maximum (VBM) positions are given with respect to the Fermi level of the substrate ($E_F$).

**9. Photoemission Spectroscopy**

While we refrained from measurements of intermediate $C_{60}$ coverage to minimize eventual sample changes due to prolonged illumination with UV light, we can yet draw a realistic picture of the energy level alignment based on prior art, as follows. Since the electron affinity of $C_{60}$ is in the range of 4.0 eV [6] to 4.9 eV (our measurements, Fig. S9A in the SI), contact with a substrate of comparable work function will result in Fermi level pinning, i.e., electron transfer from the substrate to the acceptor $C_{60}$ and thus partial filling of its LUMO manifold [7]. In our case the substrate is the perovskite film, and the electron transfer to the molecular layer will result in reduced downward surface band bending within the perovskite [8]. The accumulated electron density with the $C_{60}$ layer promotes charge carrier diffusion away from the interface, i.e., upward energy level bending within the acceptor layer occurs [7,9], in full analogy to band bending in conventional semiconductors. Note that the amount of energy level bending, as well as final work function and Fermi level position within the energy gap of the acceptor depend on the amount of transferred charge and details of the actual density of states distribution [9], which can be notably influenced by structural disorder. In addition, we stress that the Fermi level position of a 20 nm thick $C_{60}$ layer is still determined by the substrate and does not represent the intrinsic position, as the Debye length of $C_{60}$ was reported to be several 100 nm [10]. Consequently, there are three components that contribute to the work function of the 20 nm $C_{60}$ layer, i.e., reduced downward surface band bending within the perovskite, an interface dipole due to the transferred charge, and upward energy level bending within the $C_{60}$ film. Attending to the two specific examples examined here, we first observe that the work function of 0% Sr perovskites was consistently 0.15 eV (or more) higher than that of 2% Sr samples, and thus a larger amount of electron transfer to the $C_{60}$ layer when Sr is incorporated into the perovskite. This implies stronger energy level bending within the $C_{60}$ and indeed justifies the observation of higher work function, as well as wider energy spacing between Fermi level and LUMO for the $C_{60}$/2% Sr perovskite interface, compare to the Sr-free one.

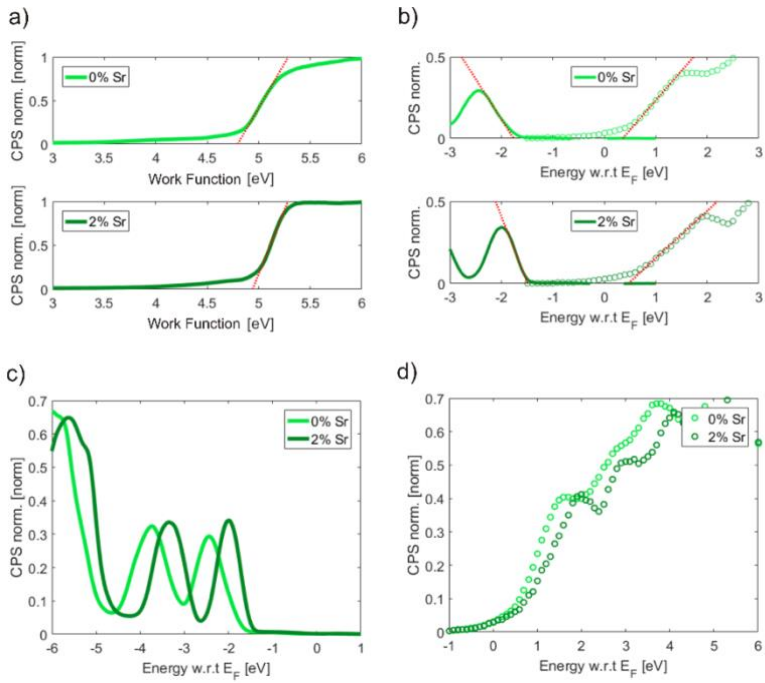

*Figure S9A:* UPS and IPES data of 20nm layer $C_{60}$ deposited on two different ITO/PTAA/perovskite samples, containing 0% and 2% Sr respectively. a) Secondary electron cut-off regions. b) Magnified valence and conduction band region near $E_F$. c) Wide range of valence band structure obtained from UPS and d) conduction band structure obtained from IPES.

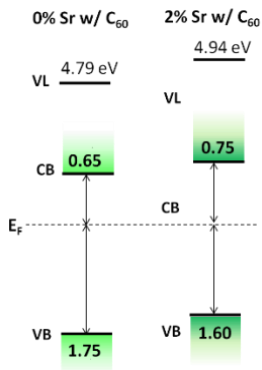

*Figure S9B:* Valance band maximum, conduction band minimum and work function energy scheme using values extrapolated from UPS/IPES (Fig. S9A) for two layers of 20nm $C_{60}$ deposited on two different PTAA/perovskite samples, containing 0% and 2% Sr respectively.

# 10. Photoluminescence Quantum Yield at Perovskite/C60 interface.

To study the importance of non-radiative recombination at the perovskite/$C_{60}$ interface, different samples with and w/o a 30 nm thick $C_{60}$ layer were subject to absolute PL measurements. This new set of measurements has been performed using a 445nm CW laser (adjusted ad 1 sun condition intensity) exciting our sample through an optical fiber connected to an integrating sphere. The PL spectra have been recorded with a Silicon CCD camera (Andor), calibrated with a Xe lamp of know spectral irradiance. The spectral photon density has been obtained from the corrected detector signal and the photon numbers of the excitation and emission obtained from numerical integration carried out in MATLAB R2017b. The PLQY values in Fig. S10 confirm how indeed the presence of $C_{60}$ drastically reduces the PL efficiency compared to the neat perovskite layer. The measurements also show that this reduction is significantly smaller for the Sr-containing perovskite, meaning that non-radiative recombination at this "unfavourable" interface has been reduced. This finding is perfectly in agreement with the increase in $V_{oc}$ observed in the device and the recombination scheme proposed for the Sr-containing samples, where the selectivity of the contact is improved due to the present of a larger bandgap and strong downward band bending, which limits the accessibility of holes to this interface. We acknowledge that the PL values measured for this new set of experiments during resubmission period are overall slightly lower than what we should expect from the high $V_{oc}$ of our devices, most probably due to the present unfavourable conditions of our gloveboxes which has affected sample preparation during resubmission period (most probably caused by the exceptionally elevated temperature and high degree of humidity recorded during this period of the year in Germany). However, the trend our results is solid and clear for multiple sets of samples.

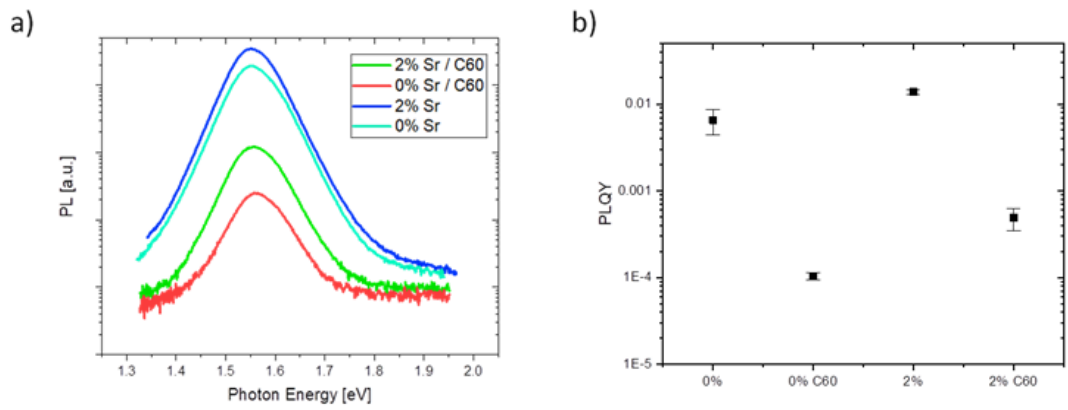

*Figure S10:* a) Photoluminescence Quantum Yield (PLQY) measurements for a neat 0% and 2% Sr sample and covered with 30nm of $C_{60}$. Averaged PLQY values, b), for the corresponding set of samples. Statistics here is calculated over 4 films for each type of sample.